\documentclass[modern]{aastex63}
\usepackage{graphicx}
\usepackage{natbib}
\usepackage{epsfig}
\usepackage{multirow}
\usepackage{amsmath}

\newcommand{\Bstar}{B_\ast}

\newcommand{\dpi}{\dot{\varpi}} 		 % apsidal rate (in eqns)
\newcommand{\dom}{\dot{\Omega}} % nodal rate (in eqns)
\newcommand{\ares}{a_{\rm res}} % resonant semimajor axis (in eqns)
\newcommand{\dd}{^\circ~{\rm d}^{-1}} % deg/day
\newcommand{\gcm}{{\rm g~cm}^{-2}} % g/cm^2
\newcommand{\fpat}{$\Omega_P$} % pattern speed (in text)
\newcommand{\ie}{{\it i.e.,}}

\newcommand{\beq}{\begin{equation}}
\newcommand{\eeq}{\end{equation}}
\newcommand{\Cassini}{{\it Cassini}}
\newcommand{\Cas}{\Cassini\ }

\newcommand{\Voyager}{{\it Voyager}}
\newcommand{\Vgr}{\Voyager\ }

\shorttitle{Saturn-driven Waves in the C Ring, v3.0}
\shortauthors{French et al.}

\begin{document}

\title{Kronoseismology V: A Panoply of Waves in Saturn's C Ring Driven by High-Order Internal Planetary Oscillations}

\author{Richard G. French}
\affil{Department of Astronomy, Wellesley College, Wellesley MA 02481}
\author{Bill Bridges}
\affil{Department of Physics, University of Idaho, Moscow, ID 83844}
\author{Matthew M. Hedman}
\affil{Department of Physics, University of Idaho, Moscow, ID 83844}
\author{Philip D. Nicholson}
\affil{Department of Astronomy, Cornell University, Ithaca NY 14853}
\author{Christopher Mankovich}
\affil{Division of Geological and Planetary Sciences, Caltech, Pasadena, CA 91125}
\author{Colleen A. McGhee-French}
\affil{Department of Astronomy, Wellesley College, Wellesley MA 02481}

\correspondingauthor{Richard G. French}
\email{rfrench@wellesley.edu}

\begin{abstract} 
Saturn's rings act as a system of innumerable test particles that are remarkably sensitive to periodic disturbances in the planet's gravitational field.
We identify 15 additional density and bending waves in Saturn's C ring driven by the planet's internal normal mode oscillations.
The collective response of the rings to Saturn's oscillations results in a host of inward-propagating density waves at outer Lindblad resonances (OLR) and outward propagating bending waves at outer vertical resonances (OVR).
In the emerging field of Kronoseismology, nearly two-dozen OLRs and OVRs have previously been identified in high-resolution radial profiles of the rings obtained from Voyager and Cassini occultation observations (see Hedman et al. (2019) Astron. J. 157:18 and references cited therein for a recent summary).
Here we apply similar wavelet techniques to extract and co-add phase-corrected waveforms from multiple Cassini VIMS stellar occultations.
Taking advantage of a highly accurate absolute radius scale for the rings (French et al. (2017) Icarus 290, 14), we are able to detect weak, high-wavenumber (up to $m$=14) waves with km-scale radial wavelengths.
From a systematic scan of the entire C ring, we report the discovery and identification of 11 new OLRs, two counterpart inner Lindblad resonances (ILR), and two new OVRs.
The close agreement of the observed resonance locations and wave rotation rates with the predictions of models of Saturn's interior (Mankovich et al. (2019) Ap. J. 871:1) suggests that all of the new waves are driven by Saturnian $f$-mode oscillations.
As classified by their spherical harmonic shapes, the modes in question range in azimuthal wavenumber from $m=8$ to 14, with associated resonance orders $\ell-m$ ranging from 0 to 8, where $\ell$ is the overall angular wavenumber of the mode.
Our suite of detections for $\ell-m=4$ is now complete from $m=8$ near the inner edge of the C ring to $m=14$ near 81,300 km. Curiously, detections with $\ell-m=2$ are less common.
These newly-identified non-sectoral ($\ell>m$) waves sample latitudinal as well as radial structure within the planet and may thus provide valuable constraints on Saturn's differential rotation. Allowing for the fact that the two ILR-type waves appear to be due to the same normal modes as two of the OLR-type waves, the 13 additional modes identified here bring to 34  the number of distinct $f$-modes suitable for constraining interior models.
\end{abstract}
\parskip 10pt

\section{Introduction}

Saturn's rings are a system of innumerable test particles that are remarkably sensitive to periodic external gravitational perturbations.
In the A and B rings, these perturbations primarily take the form of orbital resonances with the planet's satellites that drive both density and bending waves \citep{Shu84}.
The former are oscillations in the surface density and optical depth of the ring, somewhat analogous to the P-waves of terrestrial seismology, while the latter are out-of-plane oscillations more similar to terrestrial S-waves.
In the rings, however, the waves take the form of tightly-wrapped trailing spirals, and the restoring force is the rings' self-gravity rather than pressure or shear strength.
They are usually classified by their azimuthal wavenumber (or number of spiral arms) $m$ and their angular rotation rate or {\it pattern speed} \fpat\ relative to inertial space.
Density waves are driven at radial --- or Lindblad --- resonances, where the ring particles' eccentricity is excited, while
bending waves are driven at vertical resonances, where the ring particles' inclination is excited.
At most satellite resonances, the the local orbital mean motion of the ring particles is faster than the rotation rate of the perturbing potential and the resonances are referred to as inner Lindblad or inner vertical resonances, usually shortened to ILRs and IVRs.
Density waves propagate outwards from ILRs, while bending waves propagate inwards from IVRs \citep{Shu84}.

In the C ring, on the other hand, similar waves are driven not only by satellite resonances but also by periodic disturbances in the planet's gravitational field. These are due to both internal non-radial oscillations, or normal modes, as well as quasi-permanent global distortions in the planet's density distribution, sometimes referred to as tesseral gravity harmonics.
It is the former class of waves that is the subject of this paper.
Typical pattern speeds
of these oscillations, as measured in an inertial frame, are $1400-1900\dd$ and generally faster than the orbital mean motion in this region of $1100 < n < 1510\dd$.
The resonances involved are, therefore, outer Lindblad resonances (OLRs) or outer vertical resonances (OVRs). In most respects, the waves driven at these resonances are similar
to those at satellite resonances, except that density waves propagate {\it inwards} from OLRs, while bending waves propagate {\it outwards} from OVRs \citep{Shu84}.

About two-dozen density and bending waves have been described in the C ring in previous studies of radio and stellar occultation data obtained by the \Vgr\ and \Cas\ spacecraft \citep{Rosen91, Colwell09, Baillie11, HN13, HN14, French16, KronoIII, KronoIV}. Most of these have been identified as being driven by normal mode oscillations in Saturn, although several have values of \fpat\ close to the rotation rate of Saturn and therefore appear to be due to tesseral harmonics in the planet's gravity field \citep{HN14, MEM16, MEM18}.
First described in detail by \cite{Marley91} and \cite{MP93}, following an earlier suggestion by \cite{Stevenson82}, waves driven by planetary normal modes offer a new and powerful set of constraints on the interior structure and rotation rate of Saturn \citep{Fuller14a, Dederick18, CM19}. The key measurements are the mode oscillation frequencies, which may readily be deduced from the waves' pattern speeds (see Section 2).

In the limit of slow rotation, planetary normal modes have angular `shapes' that can be classified in terms of the standard spherical harmonic functions, $Y_\ell^m(\theta,\psi) \propto P_\ell^m(\cos \theta) e^{im\psi}$ with indices $0\leq \ell < \infty$ and $-\ell \leq m \leq\ell$.
Here, $\theta$ is the colatitude and $\psi$ is the longitude in a frame corotating with the planet. For example,
the internal density perturbations can be written in the form \citep{CM19}.
\beq
\delta\rho = \rho'_{\ell mn}(r)\ Y_\ell^m(\theta,\psi)\ e^{-i\sigma_{\ell mn}^c t},
\label{eq:rhop}
\eeq
\noindent where $\sigma_{\ell mn}^c$ is the oscillation frequency of the mode as seen in the corotating frame.
The index $\ell$ specifies the total number of nodal circles on the sphere, while $|m|$ is the number of such circles that pass through the planet's poles \citep{Marley91}; the remaining $\ell - |m|$ nodal circles are parallel to the equatorial plane.
Modes with even values of $\ell - m$ are symmetric about the equator and produce periodic radial gravitational perturbations in the equatorial plane; they are responsible for driving density waves in the rings.
Modes with odd values of $\ell - m$ are anti-symmetric about the equator and produce periodic vertical perturbations in the equatorial plane; they are responsible for driving bending waves in the rings.
The third index $n$ specifies the number of radial nodes in the mode's density profile between the planet's center and its surface, modes with $n=0$ being known as fundamental or $f$-modes.\footnote{
In addition to $f$-modes, self-gravitating spheres also exhibit lower-frequency buoyancy or $g$-modes and higher-frequency acoustic or $p$-modes, both with values of $n > 0$. But $g$-modes should not exist in a fully-convective planet and $p$-modes have frequencies too high to generate resonances in the rings.}
Modes with positive $m$ are prograde, \ie\ they rotate in the same direction as the planet,
while modes with negative $m$ are retrograde.
Predictions of mode frequencies for Jupiter and Saturn, including the effects of the planets' rapid rotation rates, were published by \cite{VZ81}.
We limit our attention to angular degrees $\ell\ge2$ on the basis that radial ($\ell=0$) modes cannot generate any external gravitational signature, and dipolar ($\ell=1$) $f$-modes would generate an unphysical oscillation in Saturn's center of mass \citep{CD01}. 

Theoretical arguments \citep{Marley91,MP93} suggested that the internal oscillations most likely to be responsible for driving density waves in the rings would be $f$-modes with $\ell = m$, known as {\it sectoral} modes.
Most of the initially-reported OLR-type density waves in the C ring were found to be consistent with this expectation \citep{HN13, HN14}.
For such modes, the only nodal lines are $m$ great circles passing through both poles of the planet, so that the contributions of all latitudes on the planet to the gravitational perturbation at a given longitude in the ring plane are in phase. For non-sectoral modes, some cancellation of positive and negative contributions is to be expected at the equator.
Furthermore, modes with $n>0$ will also suffer some cancellation of their external gravitational effects in the ring plane due to positive and negative density perturbations at different radii. Sectoral $f$-modes are thus likely to have the
strongest gravitational signature in the planet's equatorial plane, as well as having pattern speeds consistent with Lindblad resonances in the C ring.

The situation with bending waves is slightly different, as they are only driven by normal models with odd values of $\ell-m$. In this case, the strongest waves are expected to be due to $f$-modes with $\ell = m+1$, \ie\ modes with a single horizontal node at the equator.

Combining the results of all previously-published searches, 17 density waves have now been identified corresponding to all the sectoral $f$-modes with $2\leq m \leq 10$, as well as a few non-sectoral modes with $m = 7, 9$ and $11$ and $\ell - m = 2$ or 4.
All of the identified waves are associated with OLRs.
Two of the three density waves due to non-sectoral modes are located in the innermost part of the C ring, at radii less than 77,000~km, reflecting the higher pattern speeds of such modes compared with sectoral modes of the same $m$.
Completely unexpected, however, was the early discovery \citep{HN13} in the middle C ring of two nearby waves with $m=2$ and three with $m=3$, whereas theory predicts only one of each. Subsequent investigations turned up two additional density waves with $m=2$, one within the narrow and eccentric Maxwell ringlet at 87,530~km \citep{French16} and the other 11,000~km away in the inner C ring, at 76,434~km \citep{KronoIII}.
These additional waves remain puzzling but it seems likely that they arise from the mixing of $f$- and $g$-modes in a semi-convective region in Saturn's deep interior \citep{Fuller14b}.
Specifically, \cite{MF21} have recently identified the $m=2$ wave at 76,434~km (variously designated as B9 or W76.44) with the $\ell=2$, $n=1$ $g$-mode.

\cite{KronoIII} identified the first four known examples of bending waves driven by planetary normal modes, all at radii less than 77,000~km.
These waves correspond to modes with $m = 4, 7, 8$ and 9 and $\ell - m = 1$, 3 or 5.
All are associated with OVRs.

With the exception of wave W76.44 (which has $\Omega_P = 2169.3\dd$), and the rather enigmatic wave at 85,677~km (designated as W85.67 or wave $d$ of \cite{Rosen91}) --- which has $m=1$ and $\Omega_P = 2430.5\dd$ but has yet to be definitely identified \citep{HN14} --- all of these waves have pattern speeds in the range $1394 \dd < \Omega_P < 1880\dd$.

In the present paper, we apply the same pairwise-correlation and phase-corrected-wavelet techniques used by \cite{KronoIII} and \cite{KronoIV} in an automated search for additional weak density and bending waves in multiple stellar occultations observed by the \Cas VIMS instrument.
Taking advantage of a highly accurate absolute radius scale for the rings determined in previous studies of sharp-edged features \citep{paperIV}, we are able to detect weak, high-wavenumber (up to $m=14$) waves with km-scale wavelengths.
A systematic scan of the entire C ring has revealed 13 new density waves and 2 new bending waves whose wavenumbers and pattern speeds are consistent with their observed radial locations in the rings, and that match predictions based on the Saturn model developed by \cite{CM19}.
This model was itself chosen to fit the frequencies observed for the 14 previously-published waves with $4 \leq m \leq 11$.
An important byproduct of the model was a new estimate of Saturn's internal rotation rate of $\Omega_S = 818.13^{+2.41}_{-1.70}$~deg/day, corresponding to a period of 10~h 33.6~min, or about 6~min shorter than the standard (and IAU-approved) System III period of 10~h~39.5~min based on \Vgr SKR radio emission measurements \citep{Davies83}.

Although we have identified only one new sectoral mode (a density wave with $m = 11$ at 84,147~km) a major result of this new survey has been to increase significantly the number and variety of non-sectoral modes.
In addition to several more density waves with $\ell - m = 2$ and 4, we find one example each with $\ell - m = 6$ and 8.
Since the number of latitudinal nodal lines for a normal mode is equal to $\ell - |m|$, the frequencies of these modes should provide significant new constraints on latitudinal variations in density or rotation rate within Saturn.
Our survey has also added two more bending waves to the catalog, driven by modes with $m = 10$ and $11$ and $\ell - m = 1$ and 5, respectively.

\begin{figure}
 {\resizebox{6.0in}{!}{\includegraphics[angle=0]{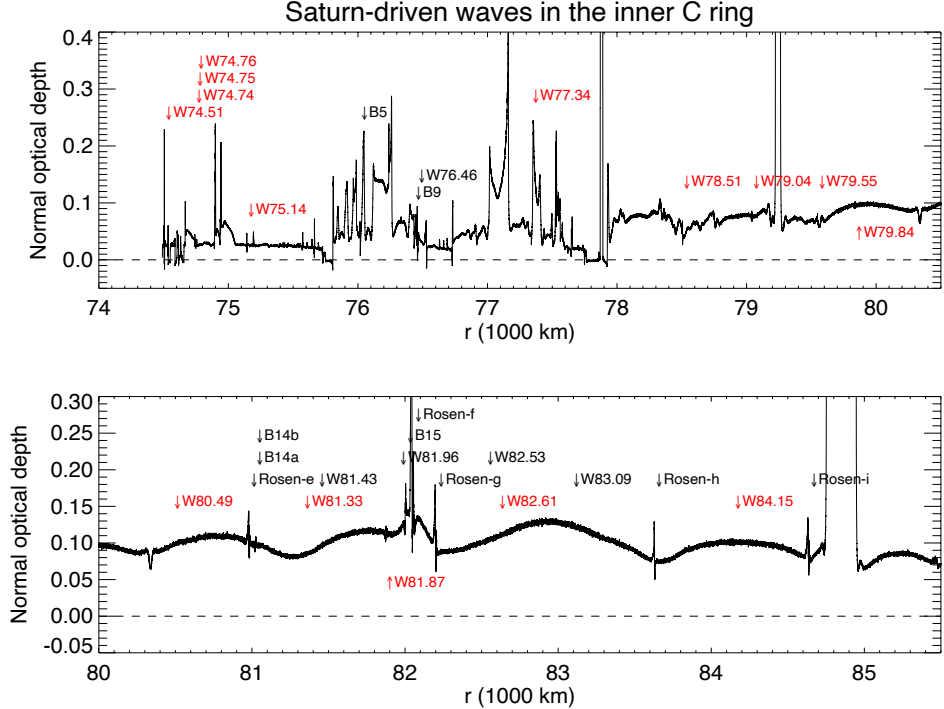}}}
\caption{Median radial profile of the normal optical depth for the VIMS stellar occultation data used for this study, spanning the inner two-thirds of the C ring.
Previously identified waves \citep{HN13,HN14,French16,KronoIII,KronoIV} are indicated by black arrows labeled with the wave name .
The 15 new identifications described in this paper are shown in red, labeled by their resonant radius in units of $10^3$~km.}
\label{fig:Cring}
\end{figure}

{\bf Figure~\ref{fig:Cring}} shows an overview of the C ring and the radial locations of all previously-detected planetary oscillation waves (labeled in black) as well as our new detections (labeled in red). The latter are labeled using the naming convention introduced by \cite{HN13}: Wxx.xx, where xx.xx is the fitted resonance radius of the wave in units of 1000 km. Whereas most of the previously-identified waves show visible structure in the optical depth profile, even at this compressed scale, nearly all of the newly-identified waves have much more subtle optical depth variations, and in some cases are not visible at all. Instead, we rely on the co-addition of data from multiple occultations to reveal the presence of these waves.

In Section 2, we collect a few useful dynamical results, while in Section 3, we summarize the observations used for our analysis, review the phase-corrected wavelet technique that underlies our search method, and develop a grading scheme for assessing the reliability of a claimed normal mode wave identification.
In Section 4, we describe our detection of 11 new density waves associated with OLRs, two with corresponding ILRs, and two new bending waves driven by OVRs, all with pattern speeds and wavenumbers consistent with waves produced by
saturnian $f$-mode oscillations.
In Section 5, we summarize our results, compare them with the predictions of current Saturn interior models, and
discuss the prospects of learning more about Saturn's internal structure and perhaps its differential rotation.

\section{Theory}

\subsection{Spiral waves in rings}

The basic theory of density and bending waves in planetary rings was laid out by \cite{Shu84}, to which the interested reader is referred for detailed derivations of the dispersion relations, propagation characteristics, and predicted amplitude variations.

The properties of the waves driven by Lindblad and vertical resonances depend upon both the perturbing forces and the properties of the rings themselves.
Adopting the useful convention that positive values of the azimuthal wavenumber $m$ correspond to an ILR or an IVR, while negative values correspond to an OLR or OVR, and denoting the resonant radius by $\ares$, for a density wave the radial wavenumber is given by the formula \citep{HN13}
\beq
|k(r)| = \frac{3(m-1)M_S}{2\pi\sigma_0\,\ares^4}\,(r-\ares),
\label{eq:disp_densitywave}
\eeq
\noindent where $M_S$ is the mass of Saturn, $\sigma_0$ is the unperturbed surface mass density of the ring and $\ares$ is the location of the exact resonance.
(The above expression does not apply for $m=1$, where an additional term involving the planet's $J_2$ is necessary; see Eq. 6 in \cite{HN13} for the full expression.)
For a bending wave the magnitude is the same but the sign is reversed due to the opposite direction of propagation away from the resonance:
\beq
|k(r)| = -\frac{3(m-1)M_S}{2\pi\sigma_0\,\ares^4}\,(r-\ares).
\label{eq:disp_bendingwave}
\eeq
\noindent (Again, a correction is necessary for the special case of $m=1$.)
Note that the waves propagate {\it outwards} from an ILR or an OVR, but {\it inwards} from an OLR or an IVR.
(On the opposite side of the resonance they are evanescent.)
The predicted linear behavior of $k(r)$, and its direction of increase, provide both a check on the reality of any putative wave, as well as way to distinguish bending from density waves.
The slope $dk/dr$ also provides an estimate of the local surface mass density, once the appropriate value of $m$ is determined, but we do not pursue this in the current study beyond a simple sanity check on $m$.

A practical consequence of these relations is that the radial wavelength of density or bending waves, $2\pi/k(r)$, decreases rapidly with increasing values of $|m|$. For data of a given radial resolution, such waves will become undetectable for sufficiently large $|m|$, or sufficiently small surface density. For the VIMS occultation data used in this work, the practical limit seems to be around $|m| \sim15$ in the middle C ring, where $\sigma_0 \simeq 5~\gcm$ and for which $2\pi/k(r) \simeq 0.7$~km at a distance of 5~km from $\ares$.

Density and bending waves are also characterized by their pattern speed \fpat\ (\ie\ their angular rotation rate relative to inertial space) and their azimuthal wavenumber $m$ (the number of arms in their spiral pattern).
The resonant radius for a Lindblad resonance is determined by the condition
\beq
(m - 1) n + \dot\varpi = m\Omega_P,
\label{eq:densitywaves}
\eeq
\noindent where $n$ is the local orbital mean motion and $\dot\varpi$ is the local apsidal precession rate due to the planet's oblateness. Resonances with $m>0$ have $\Omega_P<n$ and correspond to ILRs. Resonances with $m<0$ have $\Omega_P>n$ and correspond to OLRs.\footnote{
Note that describing OLRs or OVRs as having negative values of $m$ enables us to use a single equation to compute their radial wavenumbers and pattern speeds, but we will be associating them with prograde planetary modes with the corresponding positive wavenumber $|m|$.}
Vertical resonances satisfy the analogous resonance condition
\beq
(m - 1) n + \dot\Omega= m\Omega_P,
\label{eq:bendingwaves}
\eeq
\noindent where $n$ is again the orbital mean motion and $\dot\Omega$ is the local nodal regression rate due to the planet's oblateness. Resonances with $m>0$ have $\Omega_P<n$ and correspond to IVRs. Resonances with $m<0$ have $\Omega_P>n$ and correspond to OVRs.

In calculating the locations of Lindblad or vertical resonances in the rings, we employ standard expressions for the epicyclic ($\kappa$) and vertical ($\nu$) frequencies of near-circular, low-inclination orbits, as well as the orbital mean motion ($n$), as given by \cite{NP88} in terms of the planet's zonal gravity coefficients $J_n$.
In terms of these fundamental frequencies, we have $\dpi = n - \kappa$ and $\dom = n - \nu$. We use values of $GM_{\rm Saturn}$ and $J_2$ through $J_{12}$ as determined by \cite{Iess19} from \Cas radio science data.
As a rough rule of thumb, in the middle C ring a change of $1.0\dd$ in \fpat\ corresponds to a shift in $\ares$ of $30-35$~km, for moderate values of $m$. Typical uncertainties in our best-fitting values of \fpat\ are $0.01-0.05\dd$, corresponding to uncertainties in $\ares$ of $0.3-1.5$~km.

\subsection{Planetary normal modes}

As noted above, in a reference frame corotating with the planet, a single planetary normal mode varies with longitude and time as $e^{i(m\psi - \sigma_{\ell mn}^c t)}$. Typical $f$-mode oscillation periods $2\pi/\sigma_{\ell m0}^c$ range from 4.5~hr for $\ell=m=2$ to 60~min for $\ell = m = 20$.
In the inertial frame, the corresponding component in the planet's external potential varies with longitude and time as $e^{i|m|(\lambda - \Omega_P t)},$ where $\lambda$ refers to the inertial longitude in the plane of the rings and (see Eqns. 9 and 10 of \cite{CM19})
\beq
\Omega_P = \Omega_S + \sigma_{\ell mn}^c/m.
\label{eq:mode_freq}
\eeq
\noindent Here, $\Omega_S$ is the angular rotation rate of the planet.
We make the standard assumption that a density or bending wave driven by such a perturbation will have the same value of $|m|$ and \fpat\ as the responsible mode.

In addition to the constant frequency shift due to the change of reference frame, the mode frequency $\sigma_{\ell mn}^c$ depends on the planet's spin rate because of Coriolis and centrifugal effects.
\cite{CM19} account for rotation using perturbation theory, calculating $\sigma_{\ell mn} = m\Omega_P$ as a Taylor series in the small parameter $\Lambda = \Omega_S/\tilde{\sigma}_{\ell n}$. 
Here $\tilde{\sigma}_{\ell n}$ is the ($m$-independent) oscillation frequency for a non-rotating planet.

The predicted values of \fpat\ used in the present paper are based on theoretical calculations of $\sigma_{\ell mn}^c$ to second order in $\Lambda$.
The third-order corrections to \fpat\ are estimated to range from at most $30\dd$ for $\ell=2$ to $5\dd$ for $\ell = 15$ (see Fig.~2 in \cite{CM19}), which provides a rough estimate of the uncertainty in the predicted values of \fpat. The current uncertainties in the predicted pattern speeds are thus much greater than those in the observed values, by at least a factor of 100. 

We note here that the predicted pattern speeds for retrograde planetary modes (\ie\ those with $m<0$) are less than $\Omega_S$ and frequently negative. They are thus unlikely to result in resonances in the main rings, except possibly in the outer A ring.

\section{Observations and Methods}

Our search for Saturn-driven density waves in the C ring makes use of observations and analytical methods that have been described extensively in previous publications in this series.
We briefly describe our algorithms and methods in Sections 3.1 and 3.2, and our systematic search of the C ring in Sections 3.3 and 3.4.

As in \cite{HN13,HN14}, \cite{French16}, and \cite{KronoIV}, all of the data used in the present study come from stellar occultations observed by the Visual and Infrared Mapping Spectrometer (VIMS) instrument on Cassini \citep{Brown04}.
For our detailed analysis of wave candidates, we primarily use a subset of the data listed in Table~1 of \cite{KronoIV}, omitting events that do not cover the radial range of a particular wave, have data gaps within two km of the associated resonance radius in the propagation direction of the wave, or are contaminated by an atmospheric occultation.

\subsection{Wave identification algorithms}

In identifying candidate planet-driven waves, the major observational problem is to determine their azimuthal wavenumber $m$ and pattern speed \fpat.
Early identifications of the mysterious waves seen in the \Vgr\ radio occultation data \citep{Rosen91} were limited to comparing the radial locations of the waves with the resonance locations predicted using existing Saturn models and Eqns.~\ref{eq:densitywaves}, \ref{eq:bendingwaves} and \ref{eq:mode_freq} above \citep{MP93}.
But comparatively large uncertainties in the predicted values of \fpat\ made such identifications highly uncertain.
With the advent of large numbers of stellar occultations observed by \Cassini, this problem was solved by combining data from several different longitudes and times to measure the pattern speeds directly.
It was assumed that all such waves will exhibit variations in optical depth (real for density waves, apparent for bending waves) with inertial longitude $\lambda$ and time $t$ given by $e^{i |m|(\lambda - \Omega_P t)}$,
where the values of $m$ and \fpat\ are the same as those of the forcing mode.
Initial investigations employed a simple comparison of pairs of occultations, matching the observed phase difference between
the same wave in two observations $\delta\phi_{\rm obs}$ with that predicted for an assumed trial value of $m$:
\beq
\delta\phi_{\rm pred} = |m|(\delta\lambda - \Omega_P\delta t).
\label{eq:phi_diff}
\eeq
\noindent Here, $\delta\lambda$ and $\delta t$ are the differences in longitude and time between the two occultations.
(Note that this really involves only a single-parameter search, since for a wave at any particular radius either Eq.~\ref{eq:densitywaves} or \ref{eq:bendingwaves} can be used to estimate the appropriate value of \fpat\ for any specified value of $m$.
It is only necessary, therefore, to search over a very limited range of \fpat\ for each $m$.)
In practice, the phase difference $\delta\phi_{\rm obs}$ between any particular pair of occultation profiles is determined by an average of the phase difference between their complex wavelet transforms over a suitable range of radii, as the wavelet phase also rotates rapidly with radius.
This pairwise technique was employed by \cite{HN13}, \cite{HN14}, \cite{French16}, and \cite{KronoIII}.

This method fails, however, if the waves themselves are too weak to be seen in most individual occultation profiles, and thus to have their phases measured.
An improved technique was introduced by \cite{HN16} in their search for satellite waves in the B ring, also based on wavelet analyses.
In searching for a wave of known $m$ and \fpat, at a known radius, one can predict the phase for any given occultation $\phi_i = |m|(\lambda_i - \Omega_P t_i)$.
The complex wavelet transform of the relevant segment of data $\cal{W}$ can then be corrected to zero phase by forming the quantity $\cal{W}$~$e^{-i\phi_i}$.
Summing the corrected wavelets over many occultations, weighted appropriately, then reveals the previously-hidden wave, once the correct values of $m$ and \fpat\ are chosen.
(Again, once a value of $m$ is chosen, the pattern speed is effectively pre-determined by the local values of $n$ and $\dpi$ or $\dom$.)
This technique was used to find several weaker planet-driven waves in the inner and central C ring by \cite{KronoIII} and \cite{KronoIV}.
Besides improving the sensitivity of wave searches, the phase-correction technique is also capable of separating the signals from two or more overlapping waves.
As an example, \cite{KronoIV} were able to show that a rather confusing wave studied by \cite{HN14} at a radius of $\sim81,020$~km was in fact a composite of two waves, one with $m=-5$ and the other with $m=-11$.

Although either the pairwise-correlation or phase-corrected-wavelet technique can determine both the $m$-value and pattern speed of a wave, they cannot determine the value of $\ell$ for the mode responsible.
To date, the only way to estimate $\ell$ is to compare the observed value of \fpat\ with the predictions of interior models.
For fairly small values of $m$ and $\ell$, the frequencies are well-separated and in most cases there is a unique $\ell$ corresponding to the observed pattern speed.

\subsection{Phase-corrected wavelet technique}

Our primary search technique for planetary normal mode waves makes use of the phase-corrected-wavelet (PCW) algorithm outlined above and described in detail by \cite{HN16}.
In this subsection, we briefly summarize our implementation of the technique.
We begin by solving for the ring plane intercept time, radius and longitude as functions of the spacecraft event time at which the stellar signal was recorded by the VIMS instrument.
For this, we use our current solution for the Saturn ring plane geometry, updated slightly from that given by \cite{paperIV}.
This includes corrections to the nominal occultation geometry computed from the reconstructed Cassini spacecraft trajectory files provided by the Cassini Navigation Team shortly after the end of the Cassini orbital tour. We estimate that the corrected radius scale has a typical absolute accuracy of about 150 m.
Additionally, we apply local corrections to the radius scale to account for perturbations by the Titan 1:0 ILR, as modeled by \cite{paperII}.
Combined, these corrections have amplitudes that can exceed 1 km, and their inclusion is essential to enable accurate estimates of the phase of a density wave with a wavelength of only a few km.
We then interpolate all of the individual radial optical depth profiles onto a uniform, high-resolution grid (typically 20~m) in absolute radius, keeping track of the inertial longitude and ring intercept time of each point.

\begin{figure}
 {\resizebox{6in}{!}{\includegraphics[angle=0]{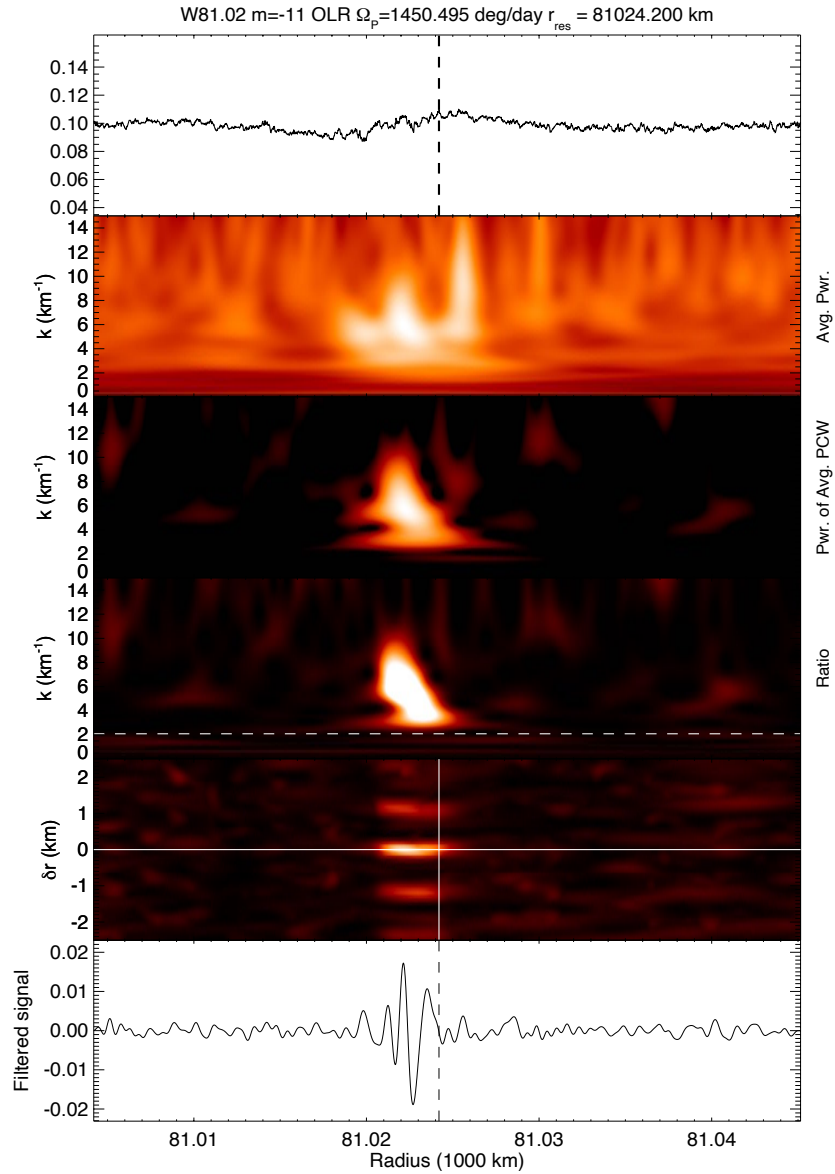}}}
\caption{Illustration of the phase-corrected-wavelet (PCW) technique for identifying waves, as applied to the W81.02 OLR-type density wave with $m=-11$ previously studied by \cite{HN14} and \cite{KronoIV}.
The first panel shows the median optical depth profile of this region; the second panel the average wavelet
power as a function of radius and radial wavenumber $k$; the third panel the average spectral power of the phase-corrected wavelet; the fourth panel the ratio of the power spectra in panels 2 and 3; the fifth panel the peak power as a function of radius and an offset in the assumed resonance radius $\delta r$; and the sixth panel the filtered wave signal reconstructed from the phase-corrected wavelet. See text for details.}
\label{fig:W81.024-2D}
\end{figure}

As described in Section 3.3, we search for density waves of a given type (ILR, OLR, IVR or OVR) and azimuthal wavenumber $m$ by solving for the average power in the phase-corrected wavelet spectrogram at each point of a high-resolution grid in resonance radius across the entire C ring.
To illustrate the effectiveness of this technique for a known wave, we will first apply this method to the W81.02 density wave studied by \cite{KronoIV}.
This wave is driven by an $m=-11$ OLR with a pattern speed $\Omega_P = 1450.50\pm0.01\dd$, but its analysis is made more challenging by the presence of an overlapping and slightly weaker density wave with $m=-5$ and $\Omega_P = 1593.63\pm0.02\dd$ (see Section 5 of \cite{HN14}).
The calculated resonance radii for the two waves are 81024.2 and 81023.2~km, respectively.
The top panel of {\bf Fig.~\ref{fig:W81.024-2D}} shows the median optical depth profile over a radial range of 40~km centered on the resonance.
This is a region of moderate optical depth ($\tau\simeq 0.1$) with no obvious wave structure in the vicinity of the resonance radius as marked by the vertical dashed line.

We first decompose each radial optical depth profile within $\pm20$ km of the assumed resonance radius $\ares$ into a series of Morlet wavelets $\cal{W}$$_i(k,r)$, and compute the combined power of the wavelet spectra, averaged over all the individual occultation profiles.
The resulting spectrogram in the second panel of Fig.~\ref{fig:W81.024-2D} shows this average wavelet power as a function of radial wavenumber $k$ and radius $r$. We see three bright regions, showing the presence of some wavelike signal in three adjacent radial ranges near the presumed resonance radius.

Next, following Eqns. 4--6 of \cite{KronoIV}, and given the observed longitude $\lambda_i$ and event time $t_i$ for each occultation --- along with the $m$-value and associated pattern speed $\Omega_P$ for the presumed resonance --- we compute the phase parameter $\phi_i=|m|[\lambda_i - \Omega_P t_i]$.
We then apply the phase correction $e^{-i\phi_i}$ to the complex wavelet for each profile to obtain the phase-corrected wavelet $\cal{W}$$_{\phi,i}$ and then sum these to form the average phase-corrected wavelet $\langle\cal{W}$$_{\phi}\rangle$ (henceforth referred to as the PCW) for the ensemble of data.\footnote{
A slight modification to this procedure is necessary for bending waves, whose apparent phase depends also on the line of sight direction to the star. For details, the interested reader is referred to Section 4 and Eq.~13 of \cite{KronoIII}.}
Note that the same trial value of $\Omega_P$ is applied to each radial location in the 40~km-wide profile, as expected for a wave forced by an external perturbation.
The power of the average PCW for $m=-11$ is shown in the third panel of Fig.~\ref{fig:W81.024-2D}.
Notice that the PCW spectrum isolates the power from the middle of the three bands of average power in the second panel, and minimizes the contributions from the other two, which tend to cancel out.
A clear wavelike signal is seen in the vicinity of 81022~km.
(The weaker wavelike signal seen at $\sim81020$~km in the second panel is likely due to the $m=-5$ wave.)

The fourth panel of Fig.~\ref{fig:W81.024-2D} shows the ratio of the average PCW spectrum to the uncorrected power spectrum (that is, the ratio of the spectrum in the third panel to that in the second panel). This is a measure of what fraction of the total wavelet power is consistent with the assumed values of $m$ and $\Omega_P$. We now see a very clear signature of $k(r)$ with a negative slope, as we would expect for an inward-propagating density wave driven at an OLR, with the wavelength decreasing away from the resonance radius.

As a test of the physical plausibility of the detection, the fifth panel shows the peak power ratio within a specified range of wavenumbers as a function of radius and pattern speed.
The latter is expressed as $\delta r$, the shift in the resonance radius relative to the nominal value assumed for this wave. For a self-consistent solution, the maximum wavelet power ratio should occur at $\delta r=0$, as seen in this example (\ie\ the best-fitting value of \fpat\ is consistent with the assumed value of $\ares$).
The range of wavenumbers included in the calculation of the maximum wavelet power ratio has a lower limit marked by the horizontal dashed line in the fourth panel --- in this case, $k=2 $ km$^{-1}$ --- and an upper limit given by the maximum value of the y axis: $k=15 $ km$^{-1}$.

Finally, in the bottom panel of Fig.~\ref{fig:W81.024-2D}, we show the reconstructed wave optical depth profile derived from the average phase-corrected wavelet, using only wavenumbers between $1.2$ km$^{-1}$ and $12$ km$^{-1}$ (\ie\ wavelengths $2\pi/k$ between 0.5 and 5 km). 
Although only indirectly related to the actual amplitude of the wave, this filtered signal reveals several cycles of an inward-propagating density wave with decreasing wavelength, exactly as expected for this OLR-type wave.
Comparing this clear wave signature to the median radial profile in the top panel shows the power of the PCW technique.

Overall, our analysis is in excellent agreement with the results shown in Fig.~3 of \cite{KronoIV} for this wave, confirming the fidelity of our independent implementation of the PCW algorithm for this paper.

\subsection{Systematic search for density waves}

We now turn to a description of our systematic search for density waves in the C ring, making use of the PCW technique described above. We begin by specifying a particular wavenumber $m$ and wave type (ILR, OLR, IVR, or OVR) and a radial range and grid spacing for our search.
For this study, we searched for waves of each of the four wave types with $|m|$ from 4 to 15 across the entire C ring (74490 -- 91983 km) at a grid resolution of 0.025 km, equivalent to a spacing in $\Omega_P$ of about $0.001\dd$ for an $m=15$ OLR in the inner C ring.
At each assumed resonance radius in the radial grid $\ares$ we computed the corresponding pattern speed $\Omega_P$ using Eq.~\ref{eq:densitywaves} or \ref{eq:bendingwaves} and applied the PCW technique as described above to determine the maximum power in the ratio spectrogram (panel 4 of Fig.~\ref{fig:W81.024-2D}) between a specified range of wavenumbers and within $\pm20$~km of $\ares$.
We also computed $\Delta r$, the difference between the radial location of the maximum power and the assumed resonance radius.
(For inward-propagating waves, such as the OLR example shown above, we expect $\Delta r$ to be negative, and for outward-propagating ILR-type waves, we expect $\Delta r$ to be positive. We will use this below as a test of the physical reality of possible wave detections.) Effectively, we perform the calculations required to produce Fig.~\ref{fig:W81.024-2D} for each of the nearly 700,000 points in our grid across the C ring, recording only the maximum power and $\Delta r$ at each point.

\begin{figure}
  {\resizebox{6.in}{!}{\includegraphics[angle=0]{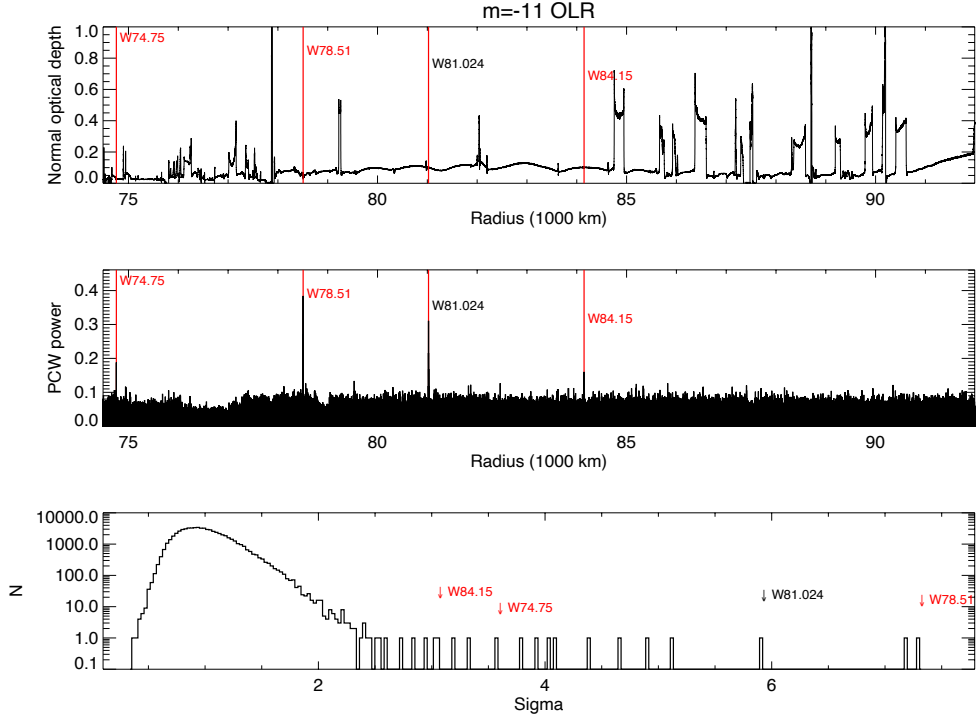}}}
\caption{Results of a search for $m=-11$ OLR-type density waves in the C ring. Top panel: C ring radial optical depth profile. Middle panel: the maximum phase-corrected wavelet power ratio as a function of resonance radius, with vertical red lines marking the four strongest detections, corresponding to three newly-detected waves (W74.75, W78.51, and W84.15) and one previously-detected wave (W81.024). Bottom panel: histogram of the ratio of the PCW power at each point in the middle panel to the local standard deviation in the vicinity of that point $\sigma$. All four of the labeled peaks are statistically significant at the $3\sigma$ level or higher. }
\label{fig:m=11OLRsearch}
\end{figure}

The results of a typical scan are shown in {\bf Fig.~\ref{fig:m=11OLRsearch}} for a search for $m=-11$ OLR-type waves.
The top panel shows the mean optical depth profile across the C ring.
The second panel shows the maximum PCW power ratio as a function of resonance radius, in arbitrary units. There are four distinct spikes in power that rise well above the background noise level, labeled by their radial locations as W74.45, W78.51, W81.024, and W84.15.
These locations are marked by vertical red lines in the upper panel as well. Only the W81.024 wave has been previously identified (it is the wave shown in Fig.~\ref{fig:W81.024-2D}); the other three are candidate waves that we describe below and identify with specific Saturn $f$-modes.
The lower panel shows a histogram of the PCW power ratio at each point in the second panel.
We denote this as $\sigma$, which is the maximum power ratio divided by the local rms variation in the same quantity.
The maximum value of $\sigma$ for each detected wave is marked by a labeled arrow.
(The other values of $\sigma>2.5$ are all associated with one of the four labeled waves, but correspond to sampled points adjacent to each peak.)

\subsection{Evaluation of candidate waves}

\begin{figure}
 {\resizebox{6in}{!}{\includegraphics[angle=0]{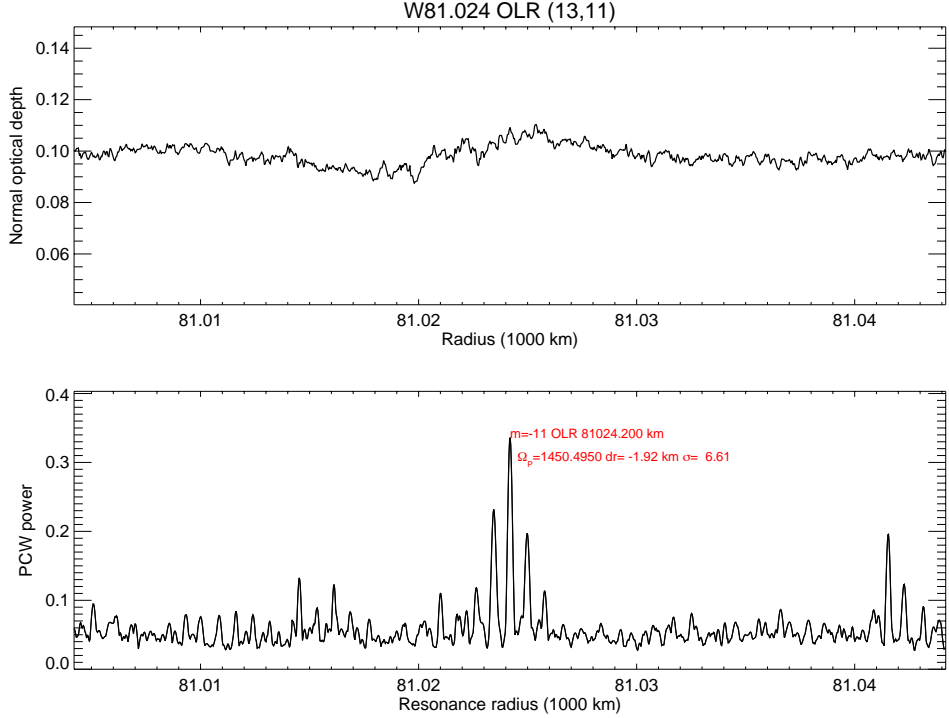}}}
\caption{A zoomed-in version of the search for $m=-11$ OLR density waves in the C ring, centered on the W81.024 wave identified by \cite{KronoIV}.
(For context, see the top two panels of Fig.~\ref{fig:m=11OLRsearch}.)
Although there is no visible wave structure in the average optical depth profile in the upper panel, the lower panel shows the sharp spike in PCW power near a resonance radius of 81024~km and a pattern speed $\Omega_P=1450.495\dd$.}
\label{fig:W81.024-1D}
\end{figure}

We conducted scans across the C ring for ILR, OLR, IVR, and OVR-type waves with values of $|m|$ between 4 and 15, producing results similar to those shown in Fig.~\ref{fig:m=11OLRsearch} for each $m$.
The next step in our search was to evaluate the wave candidates revealed in each scan.
For each tentative detection, we plotted the PCW power as a function of resonance radius in the vicinity of the nominal resonance location.
{\bf Figure~\ref{fig:W81.024-1D}} illustrates the results for the W81.024 density wave (feature B14 of \cite{Baillie11}) flagged in the automated scan with $m=-11$.
The upper panel shows the median optical depth profile while the lower panel shows the peak PCW power vs $\ares$, with the maximum corresponding to $\sigma = 6.61$ and $\Delta r = -1.92$~km.
The resonance radius is 81024.20~km and the pattern speed $\Omega_P=1450.495\dd$.
Two satellite peaks in PCW power with a spacing of about 1~km in $\ares$ are also visible.
The large value of $\sigma$ indicates the presence of a strong wave, and the negative value of $\Delta r$ is consistent with an inward-propagating wave, as expected for an OLR.

Using the resonance radius associated with the peak power, we next computed the quantities shown in the bottom three panels of Fig.~\ref{fig:W81.024-2D}: the PCW power ratio spectrum, variations in the peak power ratio with pattern speed expressed as $\delta r$, and the wave profile reconstructed from the average PCW wavelet.
(Below, when we present our results for the 15 new wave candidates, we omit the spectrograms of the average wavelet power and the average PCW power, retaining only the PCW power ratio panel.)
Notice that the fifth panel shows subsidiary horizontal bands at $\delta r \simeq \pm 1.1$~km, corresponding to the subsidiary peaks in the 1-D radial scan shown in the bottom panel of Fig.~\ref{fig:W81.024-1D}.

\begin{figure}
 {\resizebox{6in}{!}{\includegraphics[angle=0]{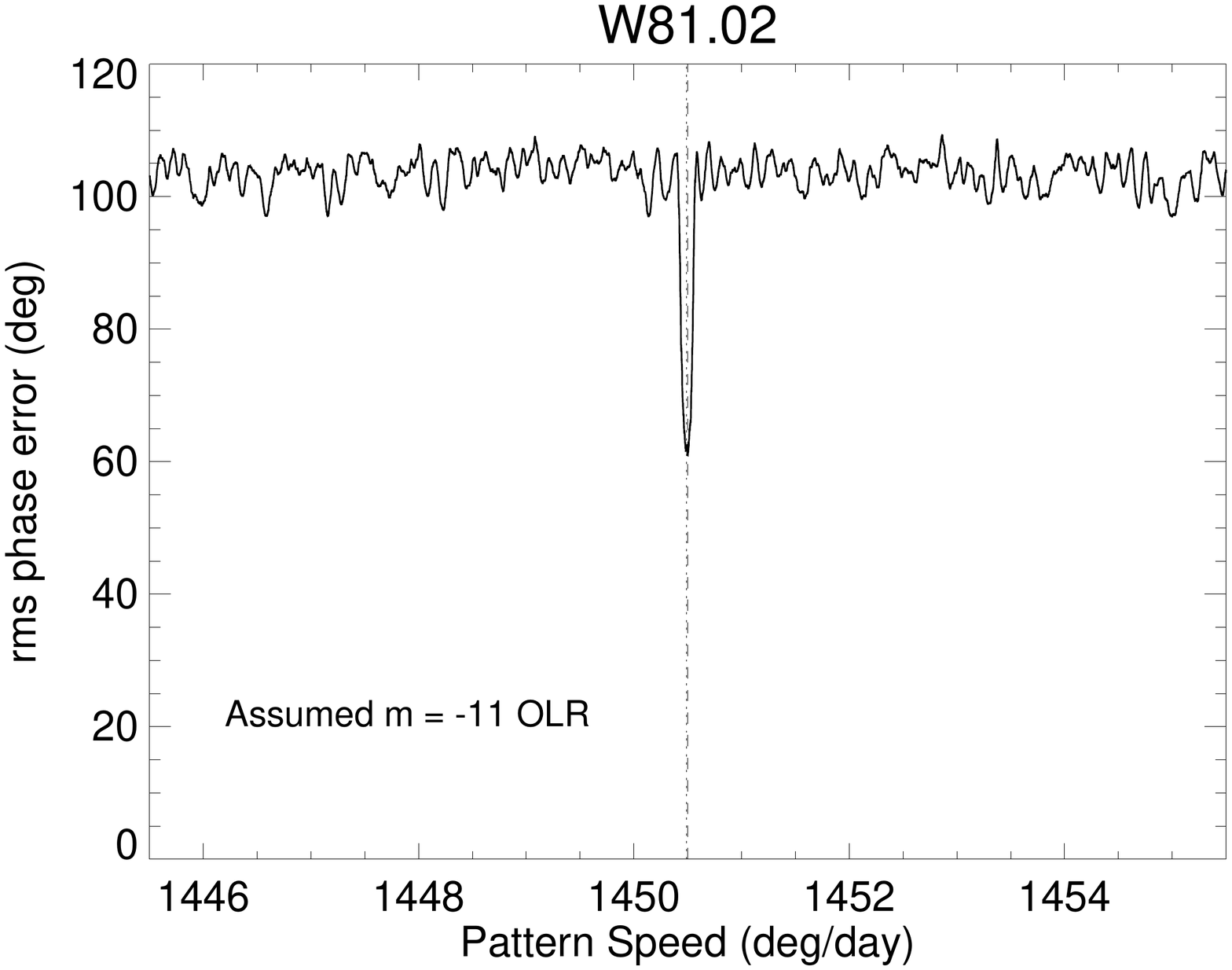}}}
\caption{Illustration of a pairwise wavelet phase comparison periodogram, showing the rms phase difference between pairs of wave profiles over a range of assumed pattern speeds for the $m=-11$ W81.02 OLR density wave first identified by \cite{HN14}.
There is a sharp minimum at a pattern speed of $\Omega_P=1450.491\dd$ (corresponding to $\ares = 81024.4 $~km), very near the pattern speed predicted for the assumed resonance radius of 81024.0~km.}
\label{fig:dphi}
\end{figure}

As a separate measure of the characteristics of each candidate wave, we applied the pairwise phase comparison method first used by \cite{HN13} to identify Saturn-driven waves, which estimates the average phase differences between pairs of wave profiles and compares these to the predicted values for a given wavenumber and pattern speed $\delta\phi_{\rm pred}$, as given by Eq.~\ref{eq:phi_diff}.
By scanning over a range of pattern speeds in the vicinity of a putative resonance, a minimum in the rms phase difference between occultation pairs can provide an independent estimate of $\Omega_P$.
For these tests, we used a much larger set of VIMS occultation data than the restricted set used for our PCW analysis, in order to increase the number of pairs of profiles and to reduce the potential for aliasing in the resulting periodogram. (This is the data set used in \cite{KronoIII} and listed in their Table 1, augmented with a few additional events from the final months of the $\Cas$ mission.)
The results for the W81.024 wave are shown in {\bf Fig.~\ref{fig:dphi}}.
We see a sharp minimum in the rms phase error (which we denote here by $\delta \phi$) at $\Omega_P=1450.491\dd$, almost identical to the values of $1450.495\dd$ found from our PCW scan and $1450.50\pm0.01\dd$ obtained by \cite{KronoIV}.

\begin{figure}
 {\resizebox{6in}{!}{\includegraphics[angle=0]{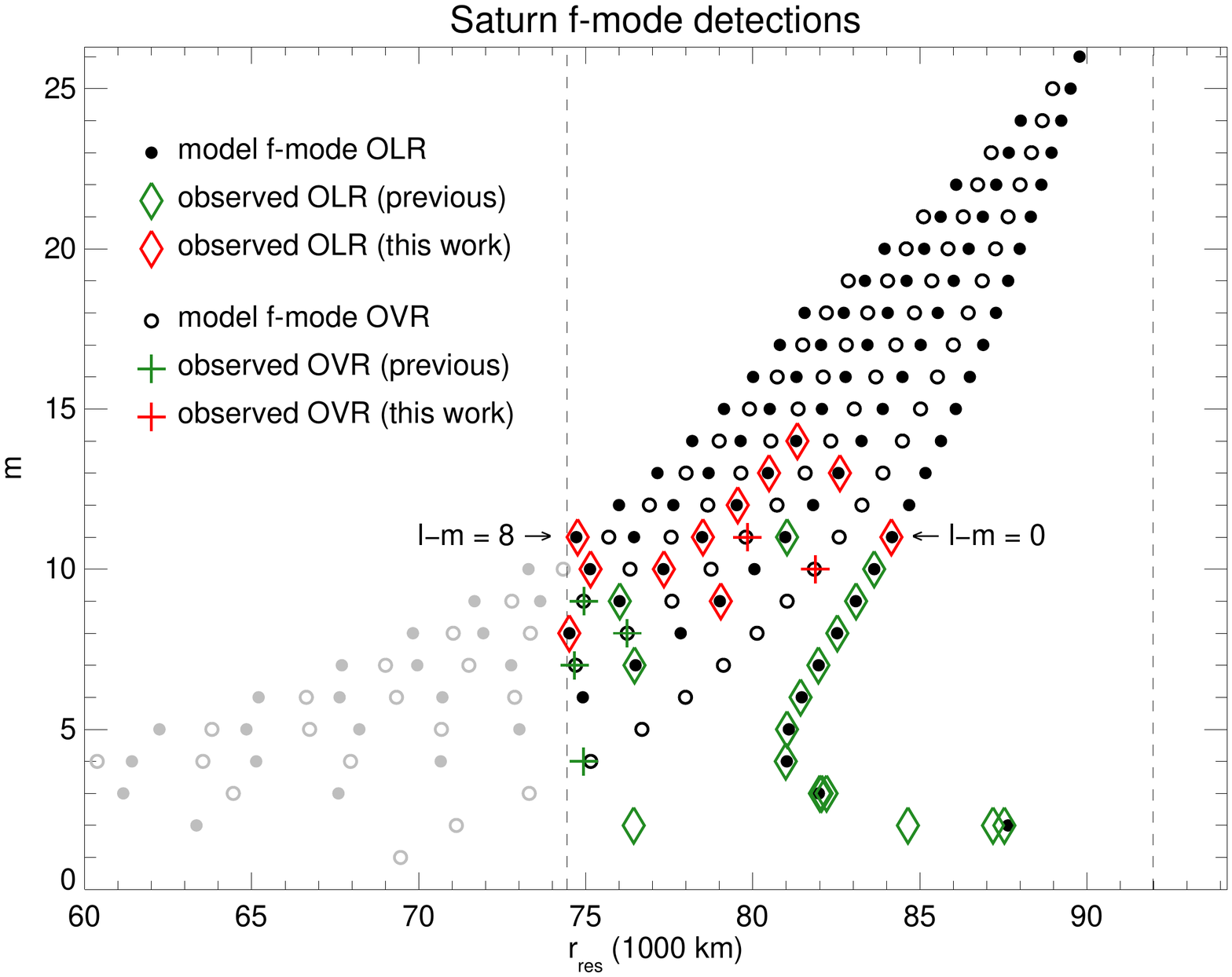}}}
\caption{Predicted and observed resonance locations of Saturnian $f$-modes in the C ring, as a function of the wavenumber $m$.
Filled circles correspond to predictions for OLR-type density waves, with even values of $\ell-m$, and open circles correspond to OVR-type bending waves, with odd values of $\ell-m$.
Two vertical dashed lines demarcate the inner and outer edges of the C ring; grayed symbols fall in the D ring. Previously published detections are shown in green (diamonds for OLRs and plus symbols for OVRs) while our new detections are shown in red. Predicted values are based on the models of \cite{CM19}.}
\label{fig:f-modes}
\end{figure}

Finally, we compared the wavenumber $|m|$ and pattern speed $\Omega_P$ of the putative wave with predictions for $f$-modes based on models of Saturn's interior by \cite{CM19}.
{\bf Figure~\ref{fig:f-modes}} shows the corresponding predicted radial locations for OLRs (filled circles, even values of $\ell-m$) and OVRs (open circles, odd values of $\ell-m$).
Because the normal mode frequencies depend strongly on the parameter $\ell$, the families of resonances with different values of $\ell-m$ are quite well-separated in such a plot.
(Note that only modes with $\ell \leq 26$ and $\ell-m \leq 8$ are shown here.)
Also shown are previously-published wave identifications in green (diamonds for OLRs; plus signs for OVRs), and our 11 new OLR waves and two new OVR waves in red.
(Not shown here are the two new ILR identifications, to avoid confusion.)
Anticipating our results below, we find that each of the waves in our final list closely matches the predicted resonance location of a specific $f$-mode, as is evident by the match between the detections and the model predictions.

Our systematic searches for ILR-, OLR-, and OVR-type waves in the C ring revealed more than two dozen candidates that warranted further investigation (no new IVR-type wave candidates were detected).
To identify the most secure of these candidates, we applied the following metrics to arrive at a numerical score for each wave:

\begin{enumerate}
\item Phase-corrected wavelet (PCW) spectrum shows significant power near the expected radial location for the wave's pattern speed: $\Delta r$ is negative for OLRs (inward-propagating waves) and positive for ILRs and OVRs (outward-propagating waves). (Score: 1 for all but the $m=11$ W79.84 wave, which used a small number of data sets.)
\item The wavenumber and the pattern speed match the expected values for the corresponding $f$-mode (to within $2\dd$, based on the \cite{CM19} Saturn internal model. (Score: 1; satisfied by all candidate waves.)
\item The median radial optical depth profile $\tau(r)$ shows a distinctive feature (not necessarily wavelike) in the vicinity of the resonance location. (Score: 1 if visible; 0 if not visible.)
\item The reconstructed waveform shows clear wavelike behavior of appropriate direction and shape. (Score: 1 for a strong wave; 0.5 for a plausible single cycle; 0 for no convincing wave structure.)
\item The RMS pairwise phase error has a recognizable minimum at the expected pattern speed. (Score: 1 for minimum phase error less than $75^\circ$; 0.5 for a weak but visible minimum at the expected location; 0 for no minimum.)
\end{enumerate}

{\bf Table~\ref{tbl:waves}} includes the 15 waves in our candidate list that have scores of 3.0 or higher, and as we describe in more detail below, there is strong evidence for each of these wave identifications.
(In the future, we will investigate the remaining ten or so candidates to assess whether more refined data processing or careful selection of data sets used for the searches will reveal them to be more secure.)
We list the waves by their normal mode identifications, first by the resonance order $\ell-m$ and then by the mode's azimuthal wavenumber $m$.\footnote{
In this context, we ignore the sign of the value of $m$ used to compute the corresponding pattern speed of the waves in Eq.~\ref{eq:densitywaves}, as this is simply a convenience to distinguish ILRs from OLRs.}
For each wave, the table includes the resonance type (ILR, OLR, or OVR), the wave ID (in the form Wxx.xx, where xx.xx is the approximate resonance radius in thousands of km), the resonance radius $\ares$ and the corresponding pattern speed \fpat, the difference $\Delta\Omega_P$ between the $f$-mode predicted pattern speed \citep{CM19} and the observed pattern speed (\ie\ predicted minus observed), the amplitude of the PCW power ratio $\sigma$ in units of standard deviations of the mean background level, the offset $\Delta r$ between the location of peak PCW power and the resonance radius, the minimum in the rms phase difference $\delta \phi$ in the pairwise wavelet periodogram, the individual scores for metrics 1 through 5 as listed above, and the total score.

In the next section, we describe the characteristics of each wave and its association with a specific Saturn normal mode, while in Section 5 we offer some thoughts about the overall population of modes identified to date.

\begin{table}
\centering
\caption{New density and bending waves in the C ring driven by Saturn $f$-modes.}
\label{tbl:waves}
\scriptsize
%\resizebox{\textwidth}{!}
{\begin{tabular}{c c r r c c c c c c c c c c c c c}
\hline

$\ell-m$ & $\ell$ & $m$ & Type & ID & $a_{\rm res}$ & $\Omega_P$ & $\Delta\Omega_P$ &
$\sigma$ & $\Delta r$ & $\delta \phi$ (rms) & \multispan{4}{\ Metric} & &Score \\
& & & & & km & deg d$^{-1}$ &deg d$^{-1}$ &
 & km & deg & 1 & 2 & 3 & 4 & 5 \\
\hline
0 & 11 & 11 & OLR & W84.15 & 84147.1 & 1369.91 & $-0.49$ & 3.1 & $-2.1$ & 82 & 1 & 1 & 0 & 0.5 & 0.5 & 3.0\\
  &  &  & ILR & W74.75& 74748.3 & 1369.93 & $-0.51$ &2.1 & $+1.2$ & 76 &1 &1& 1 & 0.5 & 0.5 & 4.0\\
1 & 11 & 10 & OVR & W81.87 & 81872.3 & 1443.31 & $+0.83$ & 2.2 & $+3.4$ & --& 1 & 1 & 1 & 1 & 0 & 4.0\\
2 & 11 & 9 & OLR & W79.04 & 79042.3 & 1533.34 &$+0.86$ & 4.3 & $-2.2$ & 49 & 1 & 1 & 1 & 1 & 1 & 5.0 \\
 & 15 & 13& OLR & W82.61 & 82607.8 & 1390.84 & $+1.20$ &2.3 & $-2.3$ & -- & 1 & 1 & 0 & 1 & 0 & 3.0 \\
 &  &  & ILR & W74.74 & 74739.9 & 1390.84 &$+1.20$ & 4.4 & $+0.6$ & -- & 1 & 1 & 1 &1 & 0 & 4.0\\
3 & 14 & 11 & OVR & W79.84 & 79835.2 & 1487.00 &$+1.26$ & 1.5 & $-0.4$ & -- & 0.5 & 1 & 1 & 0.5 & 0 & 3.0 \\%
4 & 12 & 8 & OLR & W74.51 & 74506.9 & 1697.34 & $-0.11$ &3.6 & $-1.1$ & 82 & 1 & 1 & 1 & 1 & 0.5 & 4.5\\
& 14 & 10 & OLR & W77.34 & 77338.8 & 1569.08 & $+0.48$ &3.9 & $-1.4$ & 44 & 1 & 1 & 1 & 1 & 1 & 5.0 \\
& 15 & 11 & OLR & W78.51 & 78506.8 & 1521.41& $+0.70$ &7.9 & $-1.3$ & 48 & 1 & 1 & 1 & 1& 1 & 5.0 \\
& 16 & 12 & OLR & W79.55 & 79548.9 & 1481.15& $+0.87$ &5.4 & $-1.2$ & 49 & 1 & 1 & 1 & 1 & 1 & 5.0 \\
& 17 & 13 & OLR & W80.49 & 80486.1 & 1446.65 & $+0.97$ &3.2 & $-1.6$ & 77 & 1 & 1 & 0 & 1 & 0.5 & 3.5\\
& 18 & 14 & OLR & W81.33 & 81334.3 & 1416.73 & $+1.00$ &2.3 & $-1.9$ &83 & 1 & 1 & 0 & 1 & 0.5 & 3.5 \\
6 & 16 & 10 & OLR & W75.14 & 75143.0 & 1638.98 &$+0.54$ & 2.6 & $-0.4$ & 68 &1 & 1 & 1 & 1 & 1 & 5.0\\
8 & 19 & 11 & OLR & W74.76 & 74756.6 & 1638.39 & $+1.34$ &3.8 & $-0.0$ & 68 &1 & 1 & 1 & 0.5 & 1 & 4.5 \\
\hline
\end{tabular}}
\end{table}

\section{Density and Bending Waves Driven by Saturnian Normal Modes }

To recap the mode discussion in Section 2, Saturn's internal oscillations are classified by their
spherical harmonic shapes, described by $P_\ell^m(\cos \theta) \cos(m\psi)$, where $\ell$ is overall angular wavenumber and $m$ is the azimuthal wavenumber (and equal to the number of spiral arms of the corresponding density wave).
For even values of $\ell-m$ the oscillations are N/S symmetric and can drive both ILR and OLR density waves; odd $\ell-m$ oscillations are N/S anti-symmetric and can drive bending waves.
We will organize our presentation of the newly-detected waves by the `resonance order' $\ell-m$, as listed in Table \ref{tbl:waves}, using the shorthand notation $(\ell,m)$ to identify the mode associated with each wave.
All of the waves identified in the present study have pattern speeds consistent with $f$-modes, \ie\ with values of $n=0$.

\subsection{$\ell-m = 0$}

Fundamental sectoral normal modes -- those with $n=0$ and $\ell=m$ -- have no radial nodes within the planet, and as seen in Figs.~\ref{fig:f-modes} and \ref{fig:OmegaP}, a complete set of OLR density waves driven by sectoral modes with $m$ values from 2 through 10 has been previously identified (see \cite{KronoIV} and references therein). We now extend this series to $m=11$, including the first example of an ILR/OLR pair of waves driven by a single $f$-mode.

\subsubsection{{\rm (11,11) W84.15 OLR}}

Our automated radial scan flagged W84.15 as a weak $m=-11$ OLR wave lying midway between the Rosen-h and Rosen-i wave features \citep{Rosen91,KronoIV} in a region of low optical depth ($\tau\sim 0.1$), as seen in Fig.~\ref{fig:Cring}. {\bf Figure~\ref{fig:W84.15-2D}} shows the results of our wavelet analysis of this feature. (For these and all subsequent waves, the phase-corrected wavelet figures are contained in Appendix A.) In the top panel, there is almost no hint of wavelike structure inward of the resonance radius marked by a vertical dashed line. The PCW ratio power spectrum (second panel) shows a prominent signal in the ratio power for $k\sim4 - 8$ km$^{-1}$, and the third panel shows that the maximum power occurs for $\delta r$ near zero, as expected for a dynamically self-consistent detection. The reconstructed filtered signal in the bottom panel has recognizably wavelike structure, propagating inward from the resonance radius, as expected for an OLR. The 1-D PCW scan (not shown) revealed a conspicuous peak with $\sigma=3.1$ and a negative value of $\Delta r=-2.1$ km, consistent with an inward-propagating wave. The pairwise phase periodogram had a weak minimum of $\delta\phi=82^\circ$. The pattern speed $\Omega_P=1369.91\dd$ is close to the predicted value from the Saturn interior model for $\ell=m=11$, and this identification expands the list of detected sectoral modes to span the complete set from $m=2$ to $m=11$ (Fig.~\ref{fig:f-modes}). Overall, we score W84.15 $m=-11$ OLR as 3/5.

\subsubsection{{\rm (11,11) W74.75 ILR}}
\begin{figure}
{\resizebox{6.in}{!}{\includegraphics[angle=0]{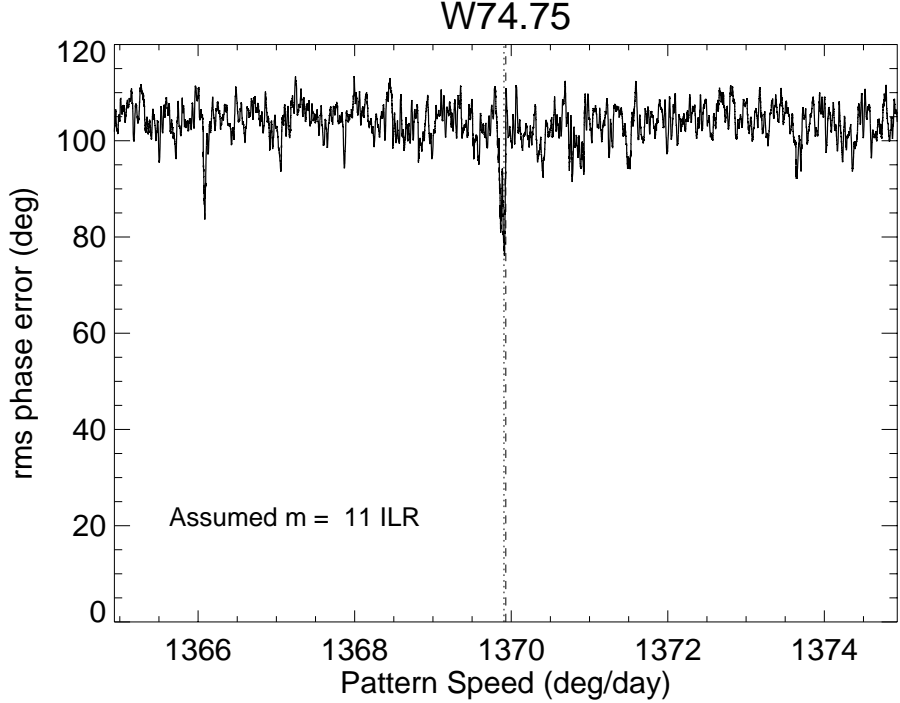}}}
\caption{The pairwise RMS phase minimization periodogram for the $m=11$ W74.75 ILR. There is a conspicuous, if somewhat noisy, minimum with an rms phase error of $76^\circ$ and a best-fitting pattern speed of $1369.9072\dd$ and a corresponding resonance radius $\ares=74749.15$ km, within a km of the value found from the PCW analysis.}
\label{fig:W74.74pairwise}
\end{figure}

The W74.75 feature (top panel of {\bf Fig.~\ref{fig:W74.75-2D}}) is the middle of a trio of low-optical depth ramp-like features near the inner edge of the C ring. From the shape of this feature, we suspected that this could be an outward-propagating wave, and from a targeted search for ILRs for $m$ from 4 to 20, we found a promising possible identification of W74.75 as an $m=11$ ILR. The 1-D scan showed a peak value of $\sigma=2.1$ and $\Delta r=+1.24$ km, and the corresponding 2D PCW ratio spectrum in the second panel of Fig.~\ref{fig:W74.75-2D} shows strong wave power rightward of the edge of the labeled resonance radius, consistent with an outward-propagating wave with the assumed pattern speed $\Omega_P=1369.93\dd$, which is close to the predicted value from the Saturn interior model for $\ell=m=11$. The third panel shows strong power near $\delta r=0$, indicating that the wave location and pattern speed are self-consistent, and the reconstructed wave profile shows a suggestion of a single cycle of an outward-propagating wave. The pairwise phase periodogram shown in {\bf Fig.~\ref{fig:W74.74pairwise}} has a convincing minimum of $\delta\phi=76^\circ$ for a pattern speed of $1369.9072\dd$, close the value found from the PCW radial scan. Collectively, these characteristics garnered a score of 4/5 for this wave.

This is the first example of an ILR/OLR wave pair driven by a single Saturn $f$-mode: $\ell=m=11$. Figure~\ref{fig:OmegaP} shows the radial locations of these matched waves on either side of the corotation frequency, connected by a dashed blue line. The pattern speeds of the two waves differ by only $0.02\dd$, comparable to the typical uncertainty of the inferred pattern speeds of the waves reported here.

\subsection{$\ell-m = 1$}
Unlike the well-populated set of observed waves driven by fundamental sectoral $f-$modes, the lowest order odd $\ell-m$ family features just a single previously-detected wave: the $m=4$ W74.94 OVR identified by \cite{KronoIII}.
This is one of only four Saturn-driven bending waves previously identified in the C ring. This paucity may indicate that N/S-asymmetric Saturn oscillations are intrinsically weaker then N/S symmetric modes, but there are observational selection effects as well. It is challenging to detect weak bending waves in occultation observations, since they are most prominent in relatively rare low-inclination events, the peak-to-trough optical depth contrast depends as well on the orientation of the wave crests to the observer, and the PCW technique does not explicitly account for the azimuthal variation in expected optical depth contrast with viewing geometry. Some of these limitations can be addressed in the future by more sophisticated analysis techniques. Nevertheless, among the several candidate OVRs identified in our searches, two were sufficiently convincing for us to report here. The first of these is the $m=-10$ W81.87 OVR.

\subsubsection{{\rm (11,10) W81.87 OVR}}
Our automated search for bending waves revealed a strong signature for the $m=-10$ W81.87 OVR near a weak-amplitude undulation in the middle C ring in a region of moderate optical depth ($\tau \sim 0.1$), shown in the top panel of {\bf Fig.~\ref{fig:W81.87-2D}}. Although there is no clear wavelike structure in the optical depth profile, the PCW power ratio spectrum in the second panel shows a strong signal with positive slope in $k(r)$, consistent with an outward-propagating wave, and the power is outward of the resonance location. The third panel, with maximum power near $\delta r=0$, confirms that the assumed pattern speed $\Omega_P$ returned by our automated search matches the assumed resonance radius. Finally, in the bottom panel, the filtered signal shows the expected structure of a wave propagating outward from the resonance radius marked by the vertical dashed line, with wavelength decreasing away from the resonance. All of these indications provide support for the reality of this as an OVR, although the corresponding phase-pair wavelet periodogram was unable to detect this wave. The measured pattern speed is very near to the predicted value for the $(11,10)$ $f$-mode (Table \ref{tbl:waves}). Overall, this bending wave has a score of 4/5, a convincing detection, marked in Fig.~\ref{fig:f-modes} with a red $+$ symbol as the second identified bending wave in the $\ell-m=1$ family.

\subsection{$\ell-m = 2$}
In contrast to the set of fundamental sectoral modes, the $\ell-m=2$ order has only one prior detection: the (13,11) W81.02 OLR that overlaps the fundamental sectoral (5,5) W81.02 OLR \citep{KronoIV}. We add two more to this list: (11,9) W79.04 OLR and (15,13) W82.61 OLR, along with its counterpart (15,13) ILR, W74.74.

\subsubsection{{\rm (11,9) W79.04 OLR}}
\begin{figure}
{\resizebox{6.in}{!}{\includegraphics[angle=0]{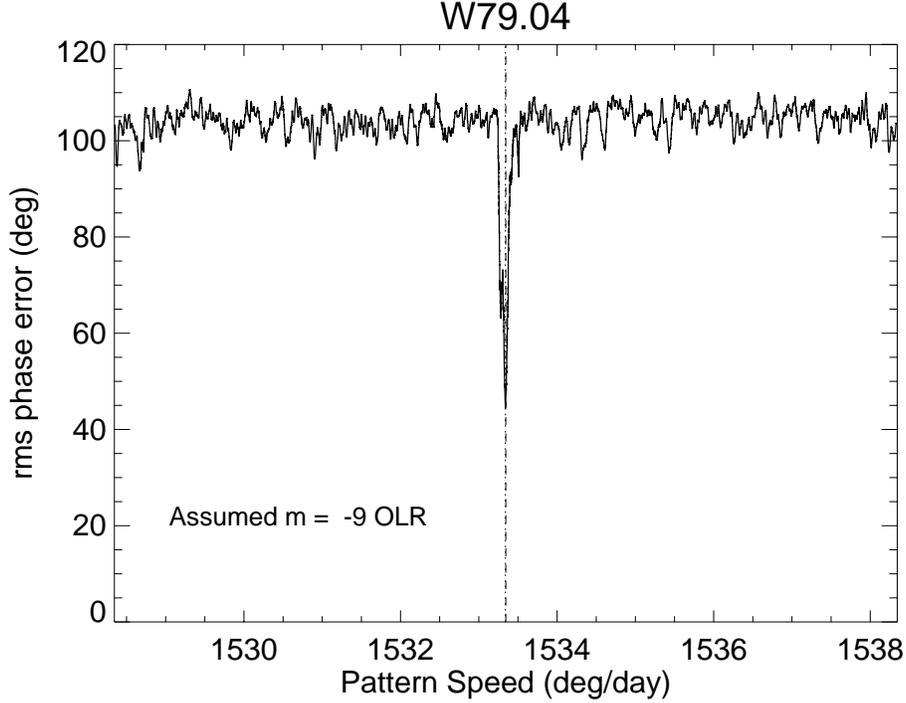}}}
\caption{Differential phase pair wavelet spectrogram for W79.04 OLR, showing a sharp minimum in rms phase error for a pattern speed $\Omega_P=1533.34^\circ$ and a resonance radius of 79042.18 km.}
\label{fig:W79.04-phasepair}
\end{figure}

The W79.04 OLR wave is located in a region of moderate optical depth $\tau\sim0.1$ (top panel of {\bf Fig.~\ref{fig:W79.04-2D}}) that lies about 200 km inward of a 40-km wide narrow ringlet (Fig.~\ref{fig:Cring}). By both quantitative and qualitative measures, this is one of the most secure identifications of our set. The 1-D PCW scan shows a very strong signature with $\sigma= 4.3$, $\Delta r=-2.24$ km, and no other significant $m=-9$ power in the vicinity. The PCW ratio power spectrum in the second panel of Fig.~\ref{fig:W79.04-2D} shows a very strong signal the characteristic negative slope of an inward-propagating wave. The power is concentrated near $\delta r=0$, and the reconstructed wave shows multiple wavecrests of an inward-propagating wave. Finally, the rms differential phase periodogram ({\bf Fig.~\ref{fig:W79.04-phasepair}}) reaches a sharp minimum of $45^\circ$ at the same pattern speed as the 1-D PCW scan: $\Omega_P=1533.34\dd$ This closely matches the model predictions for the (11,9) $f$-mode. Overall, this wave earns a solid score of 5/5.

\subsubsection{{\rm (15,13) W82.61 OLR}}

This candidate wave lies in a region of low optical depth with no visible wave structure (top panel of {\bf Fig.~\ref{fig:W82.61-2D}}), but the PCW 1-D scan for $m=-13$ OLR waves showed an isolated peak in PCW power with $\sigma=2.25$ and $\Delta r=-2.28$ km, meriting a closer look. The 2-D PCW ratio (second panel of Fig.~\ref{fig:W82.61-2D}) shows a broad negatively sloped band, with maximum power near $\delta r =0$ and a hint of an inward-propagating wave in the reconstructed profile. The observed pattern speed $\Omega_p=1390.84\dd$ is close to the predicted value for the (15,13) $f$-mode, although there is no hint of a minimum in the rms phase error periodogram in this vicinity. On balance, we regard this as a secure identification with a score of 3/5. Our confidence is strengthened by the detection of a counterpart ILR with the same pattern speed, which we describe next.

\subsubsection{{\rm (15,13) W74.74 ILR }}  
The next candidate wave in the $\ell-m=2$ series is W74.74, the innermost of three wedge-shaped features centered near 74750 km (see first panel of {\bf Fig.~\ref{fig:W74.74-2D}}), and is morphologically very similar to W74.75 ILR (Fig.~\ref{fig:W74.75-2D}).
Our automated PCW scan across the C ring gave hints of a signal near 74740 km for an $m=13$ ILR. We then ran a targeted scan at higher resolution in assumed resonance radius. We also restricted the averaged region of the PCW spectrogram to radial wavelengths below 1 km, to filter out contaminating low-frequency signals and to base our identification on sub-km wavelengths typical of high wavenumber waves in regions of low optical depth. The resulting 1-D scan in PCW ratio power showed a conspicuous signal with $\Delta r =0.64$ km and $\sigma = 4.4$, and the corresponding 2-D PCW ratio spectrum in {\bf Fig.~\ref{fig:W74.74-2D}} shows strong wave power outward of the edge of the labeled resonance radius, consistent with an outward-propagating wave with the assumed pattern speed $\Omega_P=1390.84\dd$. The third panel shows maximum power for $\delta r=0$, indicating that the wave location and pattern speed are self-consistent. The reconstructed filtered signal in the bottom panel shows a wavelike feature outward of the resonance radius. Overall, this is a convincing wave detection, with the only demerit being the absence of a visible minimum in the rms phase difference periodogram, resulting in a final score of 4/5.

We identify W74.74 ILR as the counterpart of W82.61 OLR, both driven by the (15,13) $f$-mode. Figure~\ref{fig:OmegaP} shows the radial locations of these matched waves on either side of the corotation pattern speed, connected by a dashed blue line. The pattern speeds of the two waves agree to better than $\pm 0.01\dd$.

\subsection{$\ell-m = 3$}
Only two $\ell-m=3$ OVR bending waves associated with $f$-modes have been previously identified: (10,7) W74.667 and (11,8) W76.24 \citep{KronoIII}. As we noted previously, bending waves are more challenging to identify in occultation data than radial density waves, and our automated search for $\ell-m = 3$ OVRs flagged only a small number of candidates. No obvious identifications emerged using our standard set of nearly forty VIMS occultations from the automated scan, prompting us to conduct a more targeted search using a smaller set of high-SNR, slow, and distant VIMS AlpOri occultations obtained near the end of the Cassini mission. These all share a low ring opening angle, and are thus well-suited to searching for high-wavenumber bending waves.

 In particular, we noticed
a conspicuous wavelike feature near 79835 km that was visible in three late AlpOri optical depth profiles . We constructed a 2D PCW spectrogram for each $m$ between 5 and 15 for all wave types (ILR, OLR, IVR, and OVR) and found a single instance where a strong signal persisted in the ratio power spectrum in the vicinity of a predicted $f$-mode: the (14,11) W79.84 OVR described below.

\subsubsection{{\rm (14,11) W79.84 OVR}}

The (14,11) W79.84 OVR wave is located in a region of moderate optical depth $(\tau\sim0.09)$ in the inner C ring. It is visible as a discrete, compact wavelike feature only in selected low-inclination VIMS occultations, suggesting that it is a vertical wave with short wavelength. The top panel of {\bf Fig.~\ref{fig:W79.84-2D}} shows the median profile from three AlpOri occultations in which the wave is clearly visible (revs 268E, 269E, and 277I). The second panel shows the PCW power ratio, with a strong and compact signature. From the full set of trial wavenumbers and wave types, only the $m=11$ OVR showed convincing PCW power in the radial vicinity of the wave. Because of the short time interval spanned by the three data sets, the wavelet sums normally used to construct the $\delta r$ panel provide no useful information about the correspondence between the pattern speed and resonance radius, and we omit this panel from this figure. The final panel shows the reconstructed waveform from the PCW analysis, indicating a narrow wave packet in the immediate vicinity of the resonance.

Although this wave differs from our other proposed detections in being based on a small number of occultations, we include it here because it closely matches the expected location of the (14,11) $f$-mode, it has clear wavelike appearance in low-inclination observations, and the reconstructed waveform has wavelike appearance. No other assumed wavenumbers $m=5$ to 15 of any wave type show as clear a PCW signature, and no other $f$-mode predictions match the wavenumber, pattern speed, and radial location of this feature. Given the small number of observations used, the pairwise phase minimization scan is severely aliased and does not provide useful information. Overall, we grade this wave as 3/5, reaching our cutoff for inclusion but with some weaknesses because of the limited data set used for the analysis.

\subsection{$\ell-m = 4$}
Prior to this work, the only reported instance of an $\ell-m = 4$ wave was the $m=9$ W76.02 OLR \citep{KronoIII}.
Here, we extend this to form a complete set of OLRs from $m=8$, the lowest wavenumber predicted to have an $f$-mode with a resonance radius that falls within the C ring -- see Fig.~\ref{fig:f-modes} -- to $m=14$. No ILR counterparts to these detections lie within the C ring -- all are in the D ring.

\subsubsection{{\rm (12,8) W74.51 OLR}}
The (12,8) W74.51 OLR is the innermost of our new wave identifications, located less than 20 km from the inner edge of the C ring. {\bf Figure~\ref{fig:W74.51-2D}} shows that W74.51 is located at a narrow gap just exterior to a narrow (10-km wide) spike in optical depth unassociated with any other known resonance. The 1-D PCW power ratio has a strong isolated signature ($\sigma=3.6$) at the resonance radius 74056.9 km and pattern speed $\Omega_P=1697.34\dd$, and the PCW ratio spectrogram in Fig.~\ref{fig:W74.51-2D} shows a strong signal with the expected negative slope in $k(r)$ for an inward-propagating OLR. The wave appears to be located on the steeply-sloping outer half of the 10-km wide feature. The pattern speed closely matches the $f$-mode prediction, with $\Delta \Omega_P=-0.11\dd$. The rms phase error periodogram (not shown) contains a clear minimum at the expected resonance location, albeit with a relatively weak signature with $\delta \phi=82^\circ$, resulting in a final score of 4.5/5.

\subsubsection{{\rm (14,10) W77.34 OLR }}
W77.34 ({\bf Fig.~\ref{fig:W77.34-2D}}) is located in a bland region of low optical depth ($\tau\sim0.03-0.05$) with a shape reminiscent of the outer feature in the trio of structures visible in the top panel of Fig.~\ref{fig:W74.74-2D}, which we will identify below as an $\ell=19, m=11$ OLR. There is a strong signal in the 1-D PCW scan with $\sigma=3.9$ and $\Delta r=-1.14$ km, and the PCW ratio spectrum in the second panel of Fig.~\ref{fig:W77.34-2D} shows a conspicuous band with a negative slope in $k(r)$, characteristic of an inward-propagating wave. The power is concentrated near $\delta r=0$, and the reconstructed waveform shows a very convincing shape inward of the resonance radius. There is deep minimum in the rms phase error periodogram (not shown) with $\delta \phi=44^\circ$. The fitted pattern speed closely matches the predicted $f$-mode frequency for $l=14, m=10$, with $\Delta \Omega_P=+0.48\dd$. This is a very secure identification with a score of 5/5.

\subsubsection{{\rm (15,11) W78.51 OLR}}
W78.51 ({\bf Fig.~\ref{fig:W78.51-2D}}) is similar in appearance to W77.34, with the resonance located near a low optical depth notch flanked by a peak in optical depth lying a few km inward. This $m=-11$ wave has a very strong signal in the PCW power plot with $\sigma=7.9$ and $\Delta r=-1.32$ km, small and negative, as expected for an OLR. The PCW ratio power spectrum in the second panel of Fig.~\ref{fig:W78.51-2D} shows a very strong signal with negative slope in $k(r)$. We adjusted the nominal resonance radius by 0.05 km to maximize the power near $\delta r = 0$, which slightly improved the reconstructed waveform shown in the bottom panel and more clearly revealed a slope in the PCW ratio power spectrogram. The rms phase error periodogram (not shown) has a sharp minimum of $\delta \phi= 48^\circ$.
Finally, the observed pattern speed $\Omega_P=1521.41\dd$ closely matched the prediction for $f$-mode $l=15, m=11$, with $\Delta \Omega_P=+0.70\dd$. This wave earned a solid score of 5/5.

\subsubsection{{\rm (16,12) W79.55 OLR}}

W79.55 is visually a twin of the W78.51 OLR wave just described. The optical depth profile is shown in the top panel of {\bf Fig.~\ref{fig:W79.55-2D}}. The automated 1-D PCW power ratio scan revealed a single strong spike with weaker alias lines, with $\sigma=5.4$ and $\Delta r=-1.24$ km, all consistent with an OLR. The PCW power spectrum supports this interpretation, with a very strong signal in ratio plot with a negative slope in $k(r)$. The resonance radius lies +74 km from the (16,12) $f-$mode prediction of Mankovich et al. The rms phase residual scan (not shown) features a sharp minimum at the expected resonance frequency $\Omega_P=1481.15\dd$, with $\delta \phi= 49\dd$. Finally, the observed pattern speed closely matched the prediction for $f$-mode $(l=16, m=12)$, with $\Delta \Omega_P=+0.87\dd$. This wave received a top score of 5/5.

\subsubsection{{\rm (17,13) W80.49 OLR}}

W80.49 lies 400 km interior to Rosen-e and Baillié B14a and B14b features (see Fig.~\ref{fig:Cring}). Our automatic search revealed a strong signature of an $m=13$ OLR at a resonance radius of 80486.1 km, with $\sigma=3.2$ and $\Delta r=-1.60$, although there is barely a hint of a wave in the optical depth profile at this location ({\bf Fig.~\ref{fig:W80.49-2D}}). The PCW ratio spectrum shows a strong band of power with negative slope in $k(r)$ and power concentrated near $\delta r=0$ km. Although the wave is not clearly visible in the optical depth profile, there is a convincing waveform in the reconstructed wavelet profile. The rms phase periodogram (not shown) has a sharp minimum near the predicted pattern speed, but just misses the cutoff of $\delta \phi<75^\circ$ to score a full point in this category. The fitted pattern speed nicely matched the prediction for $f$-mode $(l=17, m=13)$, with $\Delta \Omega_P=+0.97\dd$. We give this wave a score of 4.5/5.

\subsubsection{{\rm (18,14) W81.33 OLR}}

This candidate wave lies in a featureless low optical depth region exterior to the Rosen-e feature and Baillié features 14a and 14b. The 1-D PCW scan shows a modest detection with $\sigma=2.3$ and $\Delta r=-1.9$ km, made more convincing by the 2-D PCW ratio spectrum shown in {\bf Fig.~\ref{fig:W81.33-2D}}, where the reconstructed wave shows a plausible wavelike structure in the sensible direction for the proposed OLR. The rms phase periodogram (not shown) has a busy cluster of minima near the expected location, rather than a single sharp dip, with a minimum of $\delta \phi=83^\circ$. Overall, this is a convincing detection with $\Delta \Omega_P=+1.00\dd$, consistent with waves in Table \ref{tbl:waves}, but with some demerits in the overall score, resulting in a grade of 3.5/5.

\subsection{$\ell-m = 6$}
Our search detected just one $\ell-m = 6$ wave, and the only instance of a wave of this series to date. The (16,10) W75.14 OLR is among the most conspicuous of the new waves identified in this work.

\subsubsection{{\rm (16,10) W75.14 OLR}}

This $m=-10$ OLR candidate wave is associated with the first conspicuous feature outward of Baillié B3 and Rosen-b in the inner C ring (Fig.~\ref{fig:Cring}).
The shape of the optical depth profile shown in {\bf Fig.~\ref{fig:W75.14-2D}} closely resembles the $m=-10$ W77.34 wave
(Fig.~\ref{fig:W77.34-2D}) described previously. The 1-D maximum PCW power-ratio plot shows a sharp narrow spike with $\sigma=2.6$ and $\Delta r=-0.40$ km, consistent with an OLR at the calculated resonance radius. The 2-D PCW spectrogram (second panel of Fig.~\ref{fig:W75.14-2D}) shows a strong band of power with negative slope in $k(r)$, power concentrated near $\delta r=0$ km, and a convincing reconstructed waveform showing an inward-propagating wave. The rms phase periodogram (not shown) has a sharp dip with $\delta \phi= 68^\circ$, and the observed pattern speed closely matches that predicted for the ($l=16, m=10$) $f$-mode, with $\Delta \Omega_P=+0.54\dd$. This wave satisfies every metric and has a score of 5/5.

\subsection{$\ell-m = 8$}

Our search revealed the first detection of an $\ell-m=8$ wave: the (19,11) W74.76 OLR, located in the very inner C ring. With $\ell=19$, it represents the highest angular wavenumber of all Saturn-driven waves detected to date.

\subsubsection{{\rm (19,11) W74.76 OLR }}

Our final candidate wave is W74.76, identified by our automated search as an $m=-11$ OLR wave. It is located at the outer edge of the third of three distinctively shaped `fangs' in a relatively low optical depth smooth region between Baillié features B1 and B3 ({\bf Fig.~\ref{fig:W74.76-2D}}). The 1-D PCW power ratio shows a strong signal, with $\sigma=3.8$ and $\Delta r$ just below 0.0 km. The PCW ratio spectrogram shows a convincing strong signal over wavenumbers between $k=8-14$ km$^{-1}$, or about $0.5 - 1$ km in wavelength, consistent with this relatively high wavenumber mode, and there is a suggestion of the negative slope in $k(r)$ expected for an OLR. The maximum power is located near $\delta r=0$ km, and there is a plausible single cycle of an inward-propagating wave in the reconstructed waveform. The rms phase periodogram (not shown) has a sharp dip with $\delta \phi= 68^\circ$, and the observed pattern speed closely matches that predicted for the ($l=19, m=11$) $f$-mode, with $\Delta \Omega_P=+1.34\dd$. We give this wave a score of 4.5/5.

\section{Discussion and Conclusions}

\subsection{Summary of known waves due to saturnian normal-mode oscillations}

Prior to the present study, a total of 21 waves had been identified in Saturn's C ring as being likely due to normal mode oscillations within the planet: 17 density waves driven at OLRs and 4 bending waves driven at OVRs \citep{HN13,HN14,French16,KronoIII,KronoIV}.
These waves spanned values of the azimuthal wavenumber $m$ from 2 to 11 and $\ell-m$ from 0 to 5.
This work nearly doubles the list by adding 15 new waves: 11 more density waves driven at OLRs, 2 density waves driven at ILRs and 2 more bending waves driven at OVRs (see Table~\ref{tbl:waves} and Fig.~\ref{fig:f-modes}).
The current tally is then 28 OLRs, 2 ILRs and 6 OVRs.
Most of these waves have been identified with $f$-mode oscillations in Saturn (\ie\ modes with no radial nodes), but the existence of four density waves with $m=2$ and three with $m=3$ \citep{HN13,KronoIII} suggests that at least five waves are likely associated with mixed modes; these would arise either from fortuitous overlaps of $f$-mode and $g$-mode frequencies, or from spatial overlaps between the $f$-mode eigenfunctions and those of low order $g$-modes, depending on the individual case \citep{Fuller14b,MF21}.

\begin{figure}
{\resizebox{5in}{!}{\includegraphics[angle=0]{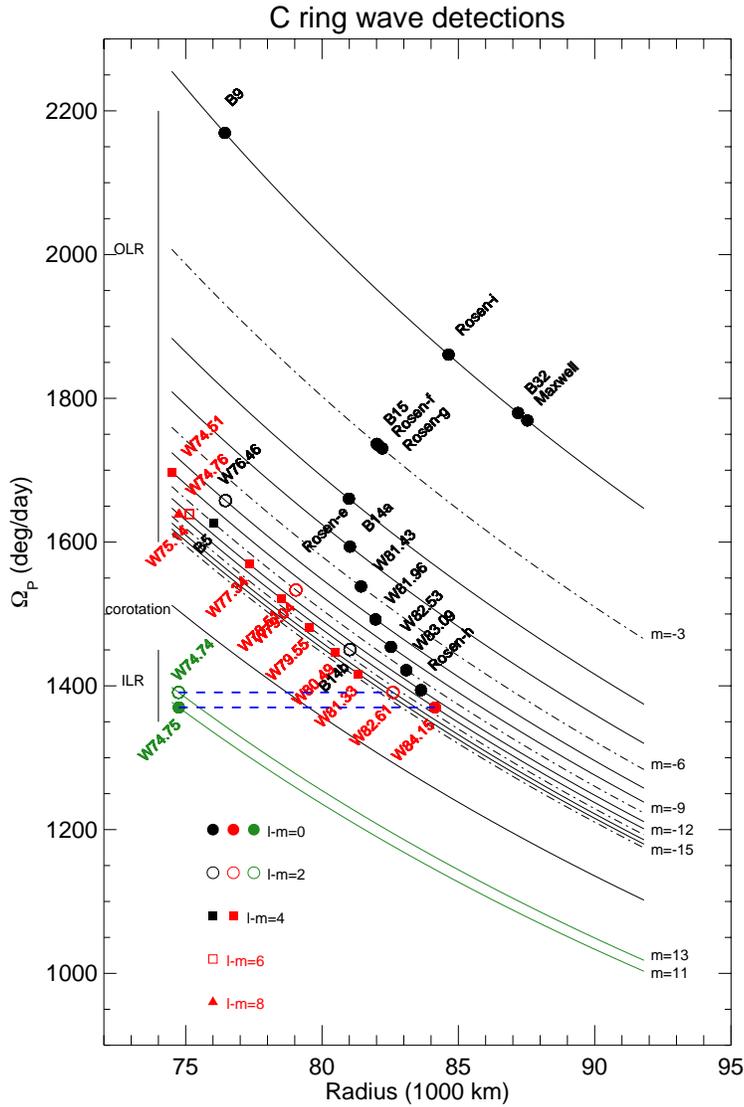}}}
\caption{Predicted patterns speeds for OLRs and ILRs in the C ring as a function of resonance radius (sloping lines), compared with wave detections (labeled symbols).
From top to bottom, OLR pattern speeds for wavenumbers from $m=-2$ to $m=-15$ are shown, followed by ILR resonances for $m=13$ and $m=11$, shown in green.
Separating these is the corotation resonance where the pattern speed is equal to the local keplerian mean motion.
Previously-detected OLR-type waves are shown as black symbols, with our new OLR detections shown in red. Green symbols mark the new ILR-type wave detections with pattern speeds and wavenumbers matching those
of two of the OLRs, as indicated by the horizontal dashed lines.
Different symbols denote the varying resonance order $\ell-m$ for the detected waves.
}
\label{fig:OmegaP}
\end{figure}

A summary of these results, complementary to that in Fig.~\ref{fig:f-modes}, is presented in {\bf Fig.~\ref{fig:OmegaP}}, which shows the calculated variation of $\Omega_P$ with radius for OLRs with $m=-2$ through $m=-15$ as well as for ILRs with $m=11$ and $m=13$, along with the measured pattern speeds for
all density waves in the C ring that have been identified with saturnian normal modes.
Previous wave detections are shown in black, while the newly-detected OLR-type waves are shown in red.
The two new ILR-type waves are shown in green.
Different symbols are used to indicate the values of $\ell-m$.
The isolated curve separating the families of OLRs and ILRs is the corotation resonance, where \fpat\ is equal to the local keplerian mean motion $n$. For OLRs, $\Omega_P > n$, and vice versa for ILRs.
Vertical resonances are omitted here for clarity, as they overlap with the Lindblad resonances at this scale.

Allowing for the fact that the two ILR-type waves appear to be due to the same normal modes as two of the OLR-type waves, the combined data set now includes a total of 34 distinct $f$-modes suitable for constraining interior models.

\subsubsection{OLRs}

By far the largest number of waves are density waves associated with outer Lindblad resonances, denoted above with negative values of $m$ and for which $\Omega_P > n$.
We categorize the 28 known waves of this type by the value of $\ell-m$ for the normal mode believed to be responsible for them.

For sectoral modes (\ie\ those with $\ell-m = 0$), we now have a complete sequence from $m=2$ to $m=11$.
For the smallest values of $m$, the resonance radius $\ares$ decreases with increasing $m$, while for $m > 5$ we see that $\ares$ increases with increasing $m$ (see Fig.~\ref{fig:f-modes}).
This is a consequence of the predicted pattern speed \fpat\ decreasing slowly with increasing $m$, while the OLR resonance condition of Eq.~\ref{eq:densitywaves} implies that $n\simeq |m|\Omega_P/(|m|+1)$. This is an increasing function of $|m|$ for small values of $|m|$.
(Note that the values of $m$ quoted here are for the normal modes, all of which are positive, rather than the negative values used in Section 4 to distinguish OLR and OVR-type waves.)

Our suite of $\ell-m=4$ detections is also unbroken, from $m$=8 very near the inner edge of the C ring at 74,507~km to $m$=14 at 81,334~km. In this case, $\ares$ increases monotonically with increasing $m$, due to the decreasing pattern speed.

Curiously, our detections of $\ell-m=2$ waves are less common, with waves being identified so far only for $m = 7$, 9, 11 and 13 (see Fig.~\ref{fig:f-modes}), ranging from 76,460~km to 82,608~km.
Also predicted to fall within the C ring are waves with $m=6$, 8, 10 and 12, none of which returned a convincing detection from our automated search.
The remaining OLR-type density waves include W75.14 with $\ell-m=6$ and $m=10$ and W74.76 with $\ell-m=8$ and $m=11$. Each of these is the innermost OLR of its order predicted to fall within the C ring.  Possibly the reason for this is that the strength of a mode's gravitational potential perturbation falls off as $r^{-(\ell+1)}$. Hence, modes with smaller resonance radii will usually have larger potential perturbations, especially for high-$\ell$ modes. 

In terms of the parameter $\ell$, we have now identified OLRs due to normal modes with all values from $\ell=2$ to $\ell=19$, including five cases with multiple values of $m$.

The limiting factor in identifying additional density waves due to planetary normal modes is probably the very short wavelengths associated with larger values of $m$, combined with the low surface mass density in the C ring (cf. Eq.~\ref{eq:disp_densitywave}).

\subsubsection{ILRs}

Density waves are also associated with inner Lindblad resonances, denoted above with positive values of $m$ and for which $\Omega_P < n$.
Most saturnian normal modes have pattern speeds too large for their ILRs to fall within the C ring.
For values of $m\geq11$, however, ILRs do begin to appear in the inner part of the C ring, moving outwards with increasing $m$. The two ILR-type density waves identified in this study are in the innermost part of the C ring, interior to 75,000~km, and correspond to $m=11$ and $m=13$.
Wave W74.75 has an $m$-value and pattern speed that match that of the OLR-type wave W84.15, and both are identified above with the $\ell=m=11$ sectoral $f$-mode.
Wave W74.74 has an $m$-value and pattern speed that match that of the OLR-type wave W82.61, and both are identified above with the ($\ell=15$, $m=13$) $f$-mode.

Dashed horizontal lines in Fig.~\ref{fig:OmegaP} connect these two ILR/OLR resonance pairs with the same wavenumber $|m|$ and pattern speed.
It is likely that additional ILR-type density waves due to modes with larger values of $m$ exist further out in the C ring, making them attractive targets for a more concerted search, but they may be difficult to identify due to their very short wavelengths (cf. Eq.~\ref{eq:disp_densitywave}).

The separation in semi-major axis between an ILR and its corresponding OLR is given approximately by
\beq
a_{\rm ILR} - a_{\rm OLR} \simeq -\frac{4m\Omega_P\,\bar{a}}{3(m+1)(m - 1)\bar{n}},
\label{eq:ressep2}
\eeq
\noindent where $\bar{a}$ and $\bar{n}$ are their average semi-major axis and mean motion.
For large $m$, $\Omega_P \simeq \bar{n}$ and $a_{\rm ILR} - a_{\rm OLR} \simeq -4\bar{a}/3m.$

 \subsubsection{OVRs}

In general, we have been less successful in identifying bending waves due to saturnian normal modes, which are expected to be driven by $f$-modes with odd values of $\ell-m$.
A total of only six such waves has been identified to date, four of which are in the inner part of the C ring.
For $\ell-m = 1$ we have only two waves, with $m=4$ and $m=10$.
For $\ell-m = 3$ we have three waves, with $m=7$, 8 and 11, while for $\ell-m = 5$ we have a single wave with $m=9$. In each case the smallest $m$-value seen corresponds to the innermost OVR of its order predicted to fall within the C ring.

The relative paucity of bending waves is more likely to reflect the greater difficulty in identifying them in occultation profiles than any intrinsic lack of such waves in the rings. Vertical distortions in the rings produce a negligible signature in the apparent optical depth of the ring both for stars at high inclination angles $\Bstar$ and for stars at lower values of $\Bstar$ where the line-of-sight is almost parallel to the wave crests \citep{KronoIII}. This significantly reduces the number of useful occultations.
Furthermore, our current phase-correction algorithm is relatively crude for bending waves, whose apparent phase depends not only on time and inertial longitude, as in Eq.~\ref{eq:phi_diff}, but also on the longitude relative to the stellar direction.
Future improvements in this algorithm may enable us to identify additional bending waves in the existing occultation profiles.

In terms of the parameter $\ell$, we have now identified OVRs due to normal modes with $\ell=5$, 10, 11 and 14, including two cases with multiple values of $m$.

Although density and bending waves differ in their mode of generation and in their direction of propagation, their pattern speeds and dispersion relations are very similar (cf. Eqns.~\ref{eq:disp_densitywave} through \ref{eq:bendingwaves} )
%and Eqns.~\ref{eq:disp_densitywave} and \ref{eq:disp_bendingwave})
and for the purpose of constraining interior models of Saturn they are equivalent, except that $\ell-m$ is even for the former and odd for the latter \citep{MP93}.

\subsection{Comparison of predicted and observed pattern speeds.}

We have seen already in Fig.~\ref{fig:f-modes} that the observed resonance radii for the identified $f$-modes generally match the predicted values to within a few tens of km, corresponding to a match in pattern speeds of $\sim1\dd$. 
This is generally more than sufficient to permit an unambiguous identification of the mode responsible for driving a given wave, especially as the data also permit a determination of the azimuthal wavenumber $m$,
although an exception must be made for the four $m=2$ and three $m=3$ density waves identified by \cite{HN13, French16} and \cite{KronoIII}.
Among the $m=2$ waves, current models suggest that the wave in the Maxwell ringlet at 87,530~km is driven by the $\ell=2$, $n=2$ $g$-mode and the wave at 84,643~km is driven by the $\ell=2$ $f$-mode. However, even these two rather well-separated modes have mixed $f/g$ character because their eigenfunctions overlap inside Saturn, and thus the labels $g$ and $f$ applied to the two are slightly arbitrary. In this case, the mode labeled $f$ is the one with the larger gravitational potential perturbation amplitude at Saturn's surface, when all the modes are normalized by their mode energy.
The wave at 76,434~km (variously designated as B9 or W76.44) is driven by the $\ell=2$, $n=1$ $g$-mode \citep{MF21}. The remaining $m=2$ wave close to the Maxwell ringlet may be due to a fortuitous frequency overlap between the $\ell=2$, $n=2$ $g$-mode and an $\ell\gg2$, $n\gg1$ $g$-mode \citep{Fuller14b,MF21}.
The situation is less clear for the three closely-spaced waves with $m=3$, whose pattern speeds differ by less than $7\dd$ and all of which may be of mixed $f/g$ character \citep{Fuller14b}.

\begin{figure}
{\resizebox{6.in}{!}{\includegraphics[angle=0]{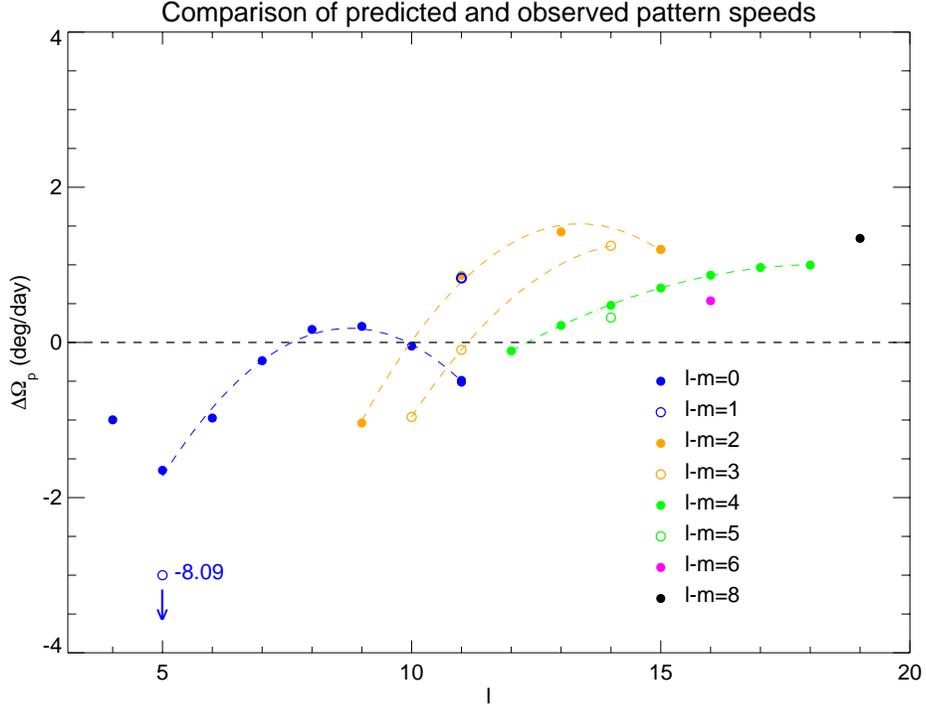}}}
\caption{Pattern speed residuals (predicted minus observed) for all waves identified with $f$-modes, plotted vs the wavenumber $\ell$ and color-coded by the order $\ell-m$.
Filled circles correspond to density waves (\ie\ even values of $\ell-m$), while open circles correspond to bending waves (odd values of $\ell-m$),
Second-order polynomials have been fitted to subsets of data with the same value of $\ell-m$ containing three or more points with $\ell \ge 4$.}
\label{fig:dOmegaP}
\end{figure}

However, the measured pattern speeds of the waves are much more precisely-determined, with typical uncertainties of $0.01-0.05\dd$.
Deviations between the predicted and observed values of \fpat\ thus provide key data with which current interior models of Saturn can be improved.
In {\bf Figure~\ref{fig:dOmegaP}} we compare the predicted minus observed pattern speeds for the full set of identified $f$-modes for $m \geq 4$, based on a representative rigidly rotating Saturn model and computational procedure used by \cite{CM19}.
We have separated modes with different values of $\ell-m$, as the non-sectoral modes sample different scales of latitudinal structure within the planet and may thus provide useful constraints on Saturn's differential rotation.

Since the observational uncertainties in \fpat\ are $<0.05\dd$ in all cases, the residuals plotted in Fig.~\ref{fig:dOmegaP} are almost entirely due to systematic errors in the predicted mode frequencies $\sigma_{\ell mn}^c$. Most of the residuals are less than $2\dd$, with the notable exception of the bending wave with $\ell=5$, $m=4$, where the discrepancy is $\sim8\dd$.
Moreover, we see that the residuals for any particular value of $\ell-m$ show a smooth variation with $\ell$ (or with $m$), strongly suggestive of a systematic error in the underlying Saturn model.
For the four cases where we have data for at least three values of $\ell$, we have fitted second-order polynomials to the residuals, indicated by the dashed lines in Fig.~\ref{fig:dOmegaP}.
In each case, the residuals are negative for small values of $\ell$, increase to a maximum, and then appear to fall again for larger values of $\ell$.
The peak residual occurs at $(9,9)$ for $\ell-m = 0$, at $(13,11)$ for $\ell-m = 2$, at $(14,11)$ for $\ell-m = 3$, and at $(18,14)$ for $\ell-m = 4$.

The residuals in Fig.~\ref{fig:dOmegaP} reflect both limitations in the accuracy of the mode frequency calculations and simplifications or incorrect assumptions underlying the Saturn interior model adopted, as well as potential errors in the equation of state for high-pressure hydrogen.
In particular, rotational corrections are included in current models only to second order in the (not very) small parameter $\Omega_S/\tilde{\sigma}_{\ell n}$, where $\tilde{\sigma}_{\ell n}$ is the oscillation frequency for a non-rotating planet \citep{CM19}.
As noted in Section 2, the third-order corrections to \fpat\ are estimated to be up to $30\dd$ for $\ell=2$, decreasing to $\sim5\dd$ for $\ell=15$, which probably provides 
% an upper limit on  %%% revised 5/03 -- PDN
some idea of the uncertainties in the predicted values of \fpat.

The neglect of differential rotation is another basic limitation of the reference model. \Cas gravity science offers persuasive evidence that Saturn's strong zonal flows penetrate some 7,000--9,000 km below Saturn's surface before yielding to a rigidly rotating interior \citep{Iess19, GK21}. Preliminary calculations show that the Coriolis forces due to such a flow would increase the inertial frame frequencies of \emph{all} prograde $f$-modes, owing to the overwhelming effect of the prograde equatorial jet \citep{MF21}. Because the velocity amplitude associated with the zonal flows is only $\sim3\%$ of Saturn's background rotation rate, this effect is rather small. Quantitative conclusions will need to await a calculation that accounts for Saturn's centrifugal distortion as well as possible higher-order effects associated with the differential rotation.

Given the complexity of the mode frequency calculations, and their dependence on the assumed rotation profile in the planet as well as the other parameters mentioned above, it is not possible to conclude anything immediately from the results shown in Fig.~\ref{fig:dOmegaP} without performing new model calculations.
Generally speaking, however, modes with larger values of $\ell$ have radial weighting functions that vary as $r^{\ell-1}$ and are thus more concentrated towards the surface of the planet. Furthermore, modes with non-zero values of $\ell-m$ will be sensitive to latitudinal variations in density, sound speed or rotation velocity.
Systematic trends in the residuals vs $\ell$ and $\ell-m$ therefore suggest corrections are necessary in the model's radial profile of sound speed and/or density, or the need to include differential rotation in the planet's outer layers.
A further complication is that $f$-modes of low $\ell$ are also sensitive to the presence of any stable (\ie\ semi-convective) layers in the planet's deep interior \citep{Fuller14b,MF21}.

\subsection{Future work}
With the expanded catalog of Saturn-driven waves, augmented by the new waves identified here, there is still no obvious pattern in the amplitudes of the detected waves, or an explanation for the apparent absence of some waves predicted to be associated with $f$-modes. Simple equipartition arguments would imply comparable amplitudes for many of the detected waves that in fact show strikingly different amplitudes. It will be helpful to quantify the strength of the detected waves more systematically and to derive upper limits on the non-detections.

A glance at Fig.~\ref{fig:OmegaP} suggests that many more high-$m$ waves are likely to exist in the outer C ring, but with radial wavelengths that are too short to detect with our current methods.
It may be possible to detect such high-wavenumber ($m\ge15$) Saturn-driven OLR waves (and their ILR counterparts in the inner and middle C ring) by concentrating on the highest-resolution VIMS occultations, augmented by selected \Cas UVIS and RSS observations.
As part of this effort, a more concerted search for OVR-type bending waves is in order, taking proper account of the geometry of each occultation to optimize their detectability.

\section{Acknowledgements}

The authors would like to thank two anonymous reviewers for their insightful comments, several of which were incorporated into the text. 
This work was supported in part by NASA CDAP grants NNX09SE66G and 80NSSC10K0890.

\section{Data availability}
All of the occultation data from which our results were obtained are publicly available from NASA's Planetary Data System Ring-Moon Systems Node at {\tt https://pds-rings.seti.org/ringocc/}.

\appendix{}
\section{Phase-corrected wavelet figures}
For convenient reference, this appendix contains the similarly-formatted phase-corrected wavelet plots for each identified wave.

\restartappendixnumbering

\begin{figure}
{\resizebox{5.5in}{!}{\includegraphics[angle=0]{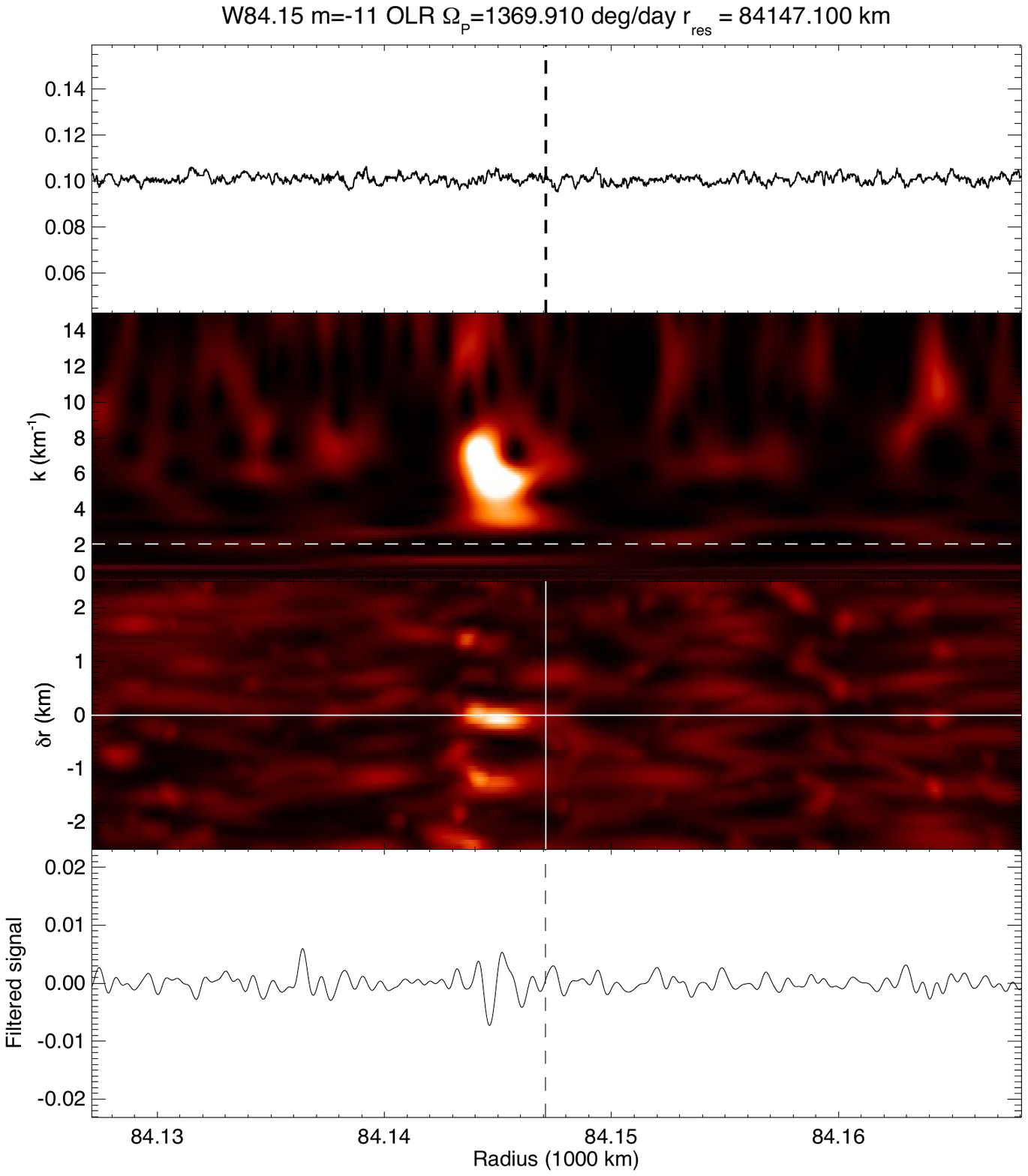}}}
\caption{Phase-corrected wavelet (PCW) results for the W84.15 $m=-11$ OLR, for $\Omega_P=1369.910\dd$ and a resonance radius $\ares=84147.1$ km. The top panel shows the optical depth profile in the vicinity of the resonance radius marked by a vertical dashed line. The second panel shows the power ratio of the PCW spectrum, with a negative slope in $k(r)$, as expected for an OLR. The third panel shows that the power corresponds to $\delta r=0$, indicating that the detected wave has the expected resonance radius for the assumed pattern speed. Finally, the bottom panel shows the reconstructed wave profile from the PCW, consistent with a wave propagating inward from the resonance radius.}
\label{fig:W84.15-2D}
\end{figure}

\begin{figure}
{\resizebox{\textwidth}{!}{\includegraphics[angle=0]{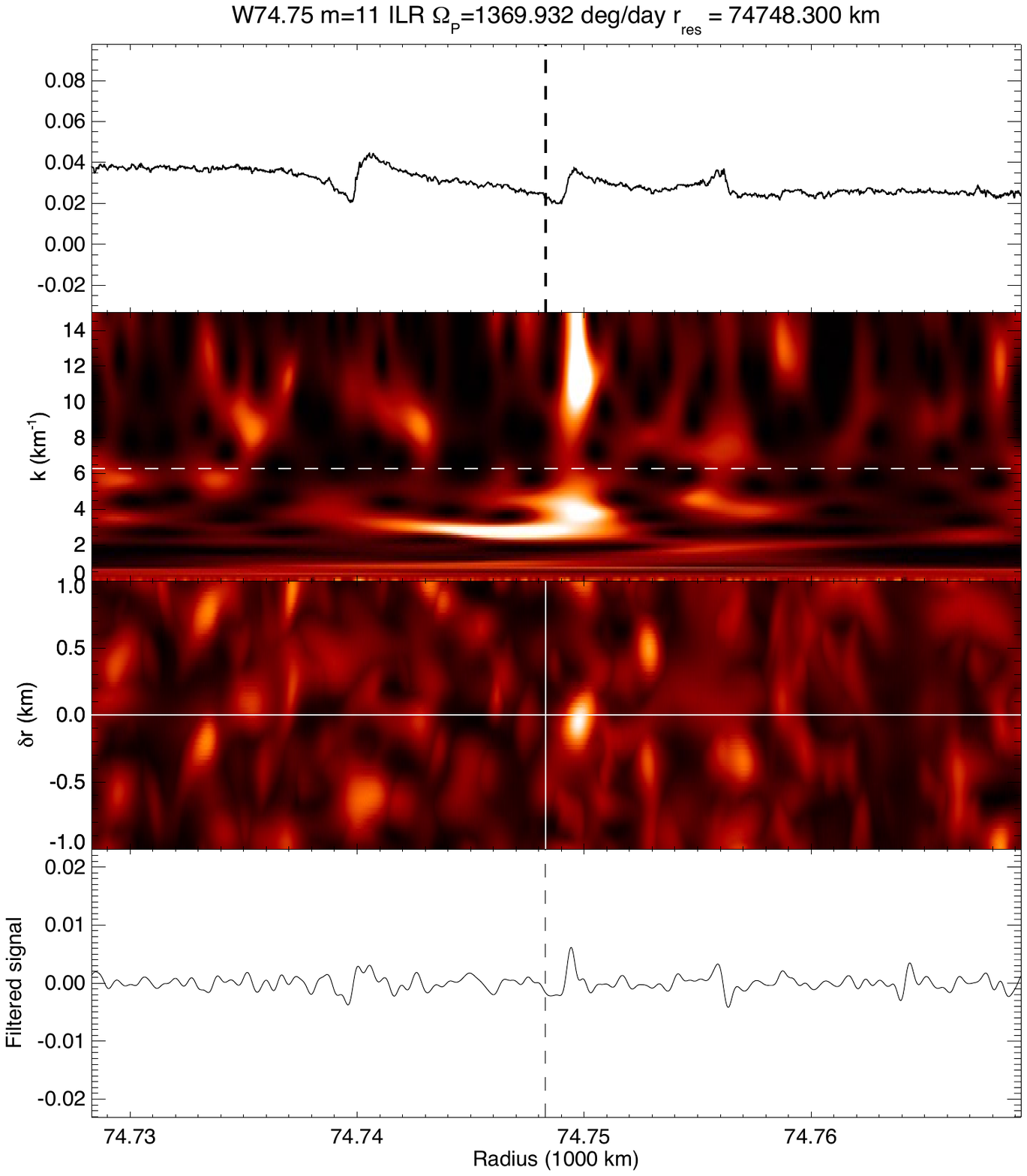}}}
\caption{Phase-corrected wavelet (PCW) results for the $m=11$ W74.75 ILR, located in the middle of a trio low optical depth features in the inner C ring (top panel). The second panel shows the usual PCW power ratio spectrogram, with significant power at high wavenumber ($k>8$), corresponding to a sub-km wavelength wave.The third panel shows that the radial location of maximum power has $\delta r=0$, indicating that the best-fitting pattern speed is consistent with the resonance radius. Finally, the bottom panel shows the suggestion of a single cycle of a weak wave outward of the resonance radius, consistent with an ILR.}
\label{fig:W74.75-2D}
\end{figure}

\begin{figure}
{\resizebox{\textwidth}{!}{\includegraphics[angle=0]{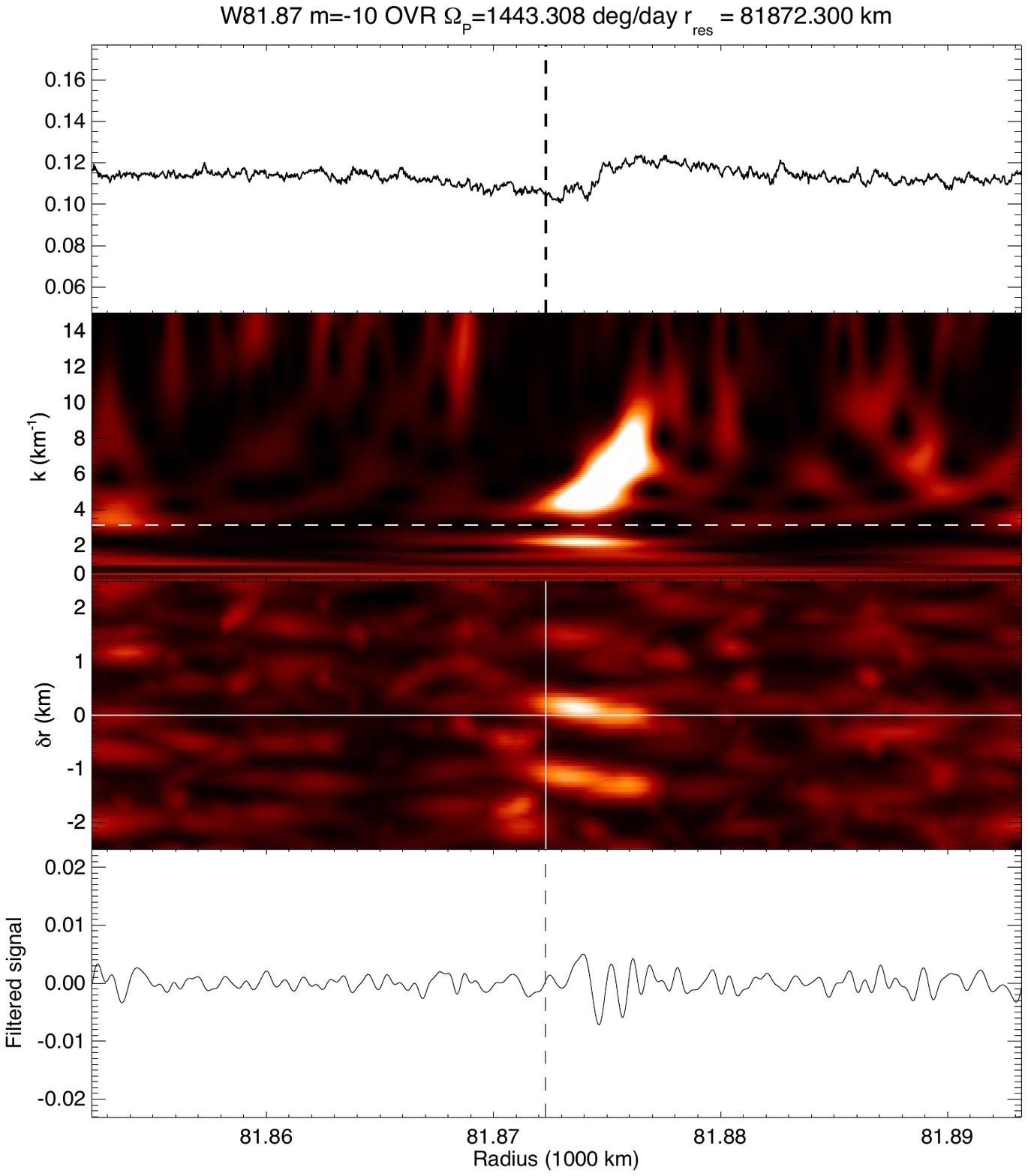}}}
\caption{Phase-corrected wavelet results for the $m=-10$ W81.87 OVR. The upper panel shows a weak undulatory feature in the optical depth profile, but no obvious wavelike structure. The second panel shows a strong feature in the PCW power ratio spectrum, with a positive slope in $k(r)$ outward of the resonance radius marked by the vertical dashed line.The third panel shows a signal near $\delta r=0$, indicating that the assumed pattern speed $\Omega_P=1443.308\dd$ is consistent with the corresponding resonance radius of 81872.3 km. The reconstructed phase-corrected wavelet filtered signal in the bottom panel has clear wavelike structure of the expected form for an OVR.}
\label{fig:W81.87-2D}
\end{figure}

\begin{figure}
{\resizebox{\textwidth}{!}{\includegraphics[angle=0]{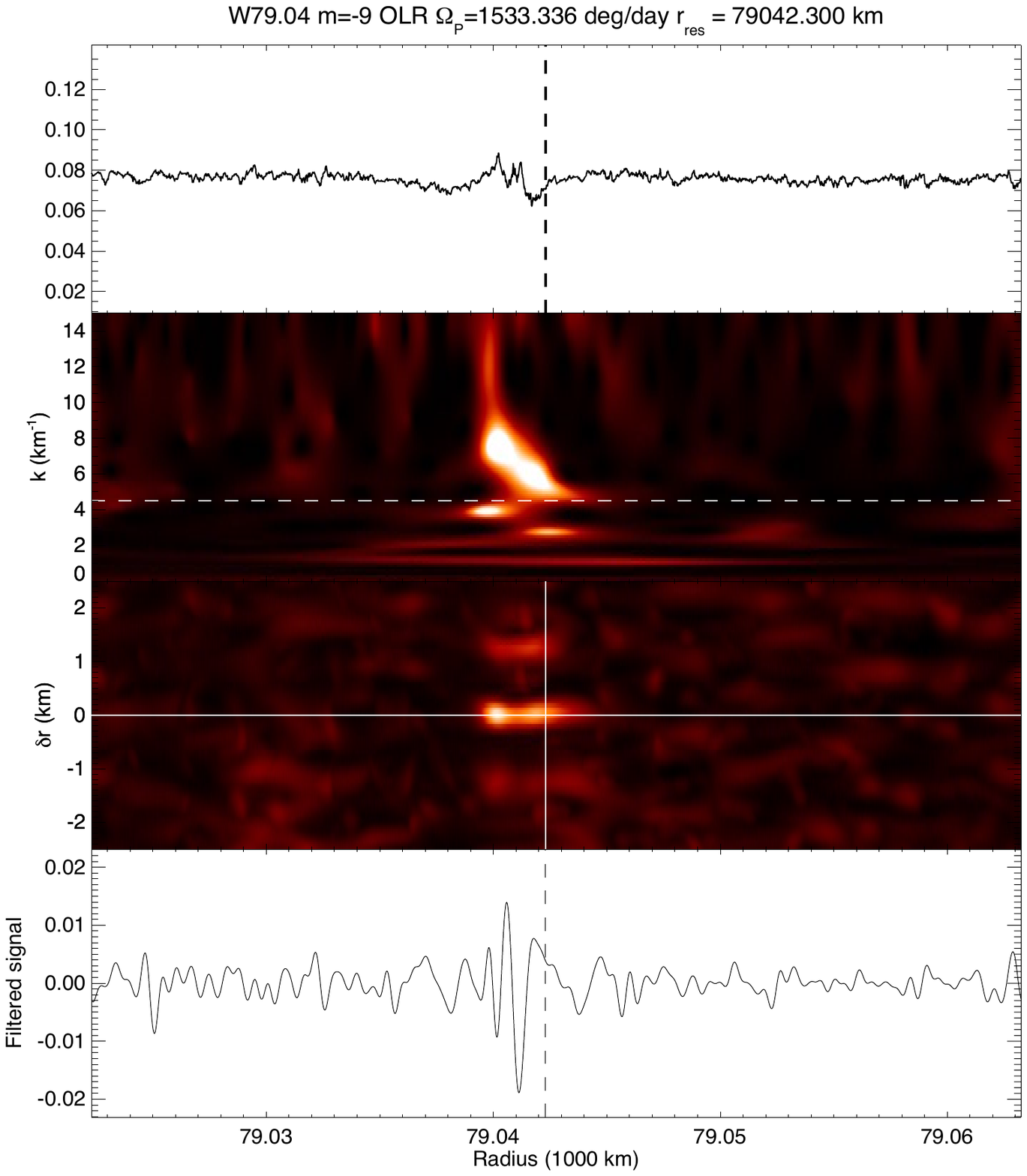}}}
\caption{Phase-corrected wavelet results for W79.04 OLR. The top panel shows a visible irregular feature just inward of the marked resonance radius. The second panel shows a strong signal in the PCW power ratio spectrum with a negative slope in $k(r)$, as expected for an inward propagating wave. The third panel shows the expected power near $\delta r=0$ indicative of a dynamically self-consistent pattern speed and resonance radius. Finally, the bottom panel shows a convincing wave structure with multiple oscillations inward of the resonance radius.}
\label{fig:W79.04-2D}
\end{figure}

\begin{figure}
{\resizebox{5.5in}{!}{\includegraphics[angle=0]{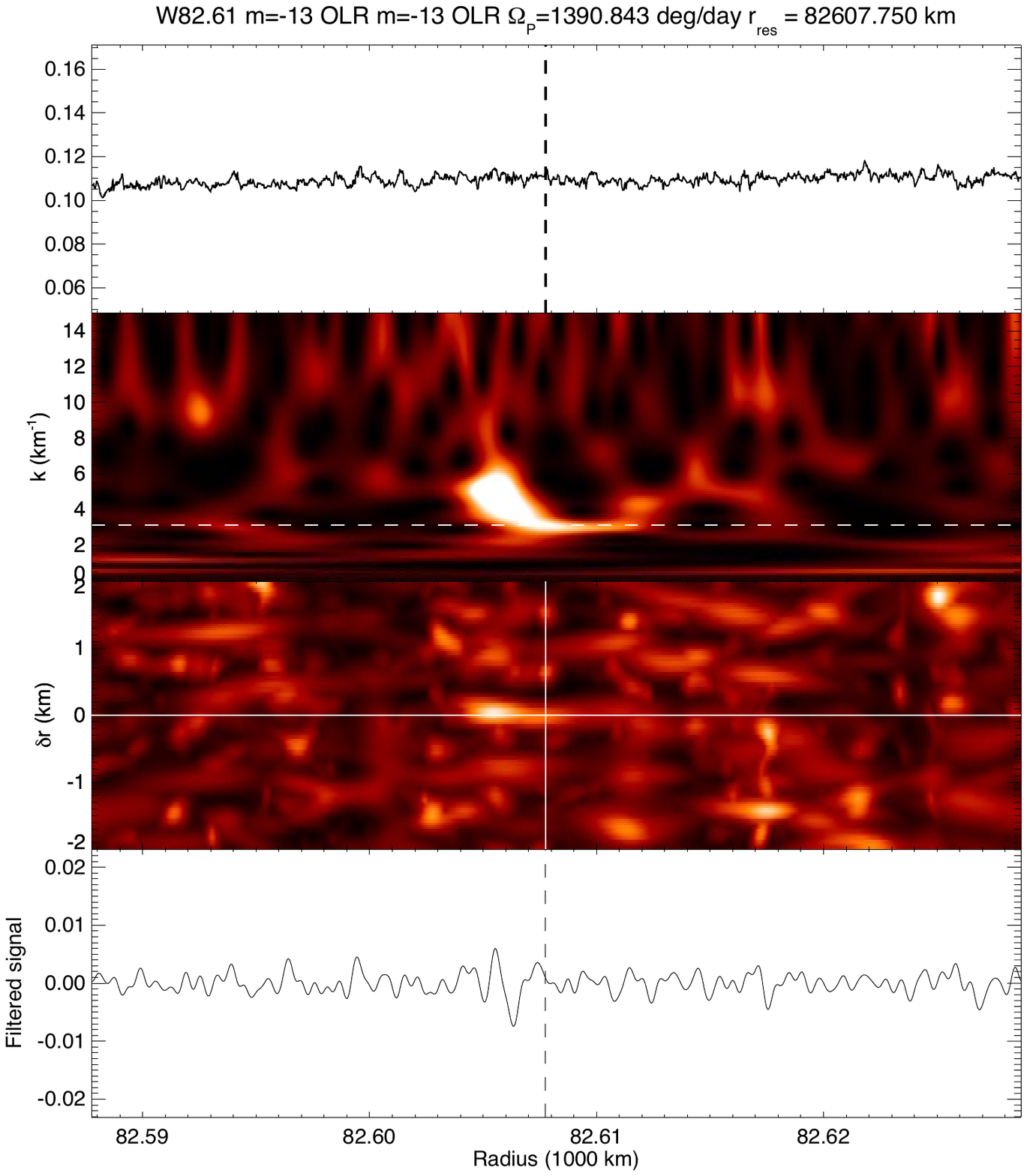}}}
\caption{Phase-corrected wavelet results for the $m=-13$ W82.61 OLR. The top panel shows a featureless region of moderate optical depth, with no hint of a wave. However, the PCW power ratio spectrum in the second panel shows a strong signal with a negative slope in $k(r)$, inward of the assumed resonance radius marked by the vertical dashed line. The third panel shows power near $\delta r=0$, indicating that the pattern speed and resonance radius are self-consistent, and the bottom panel shows the reconstructed signal from the PCW, with a weak wavelike signature inward of the resonance radius, as expected for an OLR.}
\label{fig:W82.61-2D}
\end{figure}

\begin{figure}
{\resizebox{\textwidth}{!}{\includegraphics[angle=0]{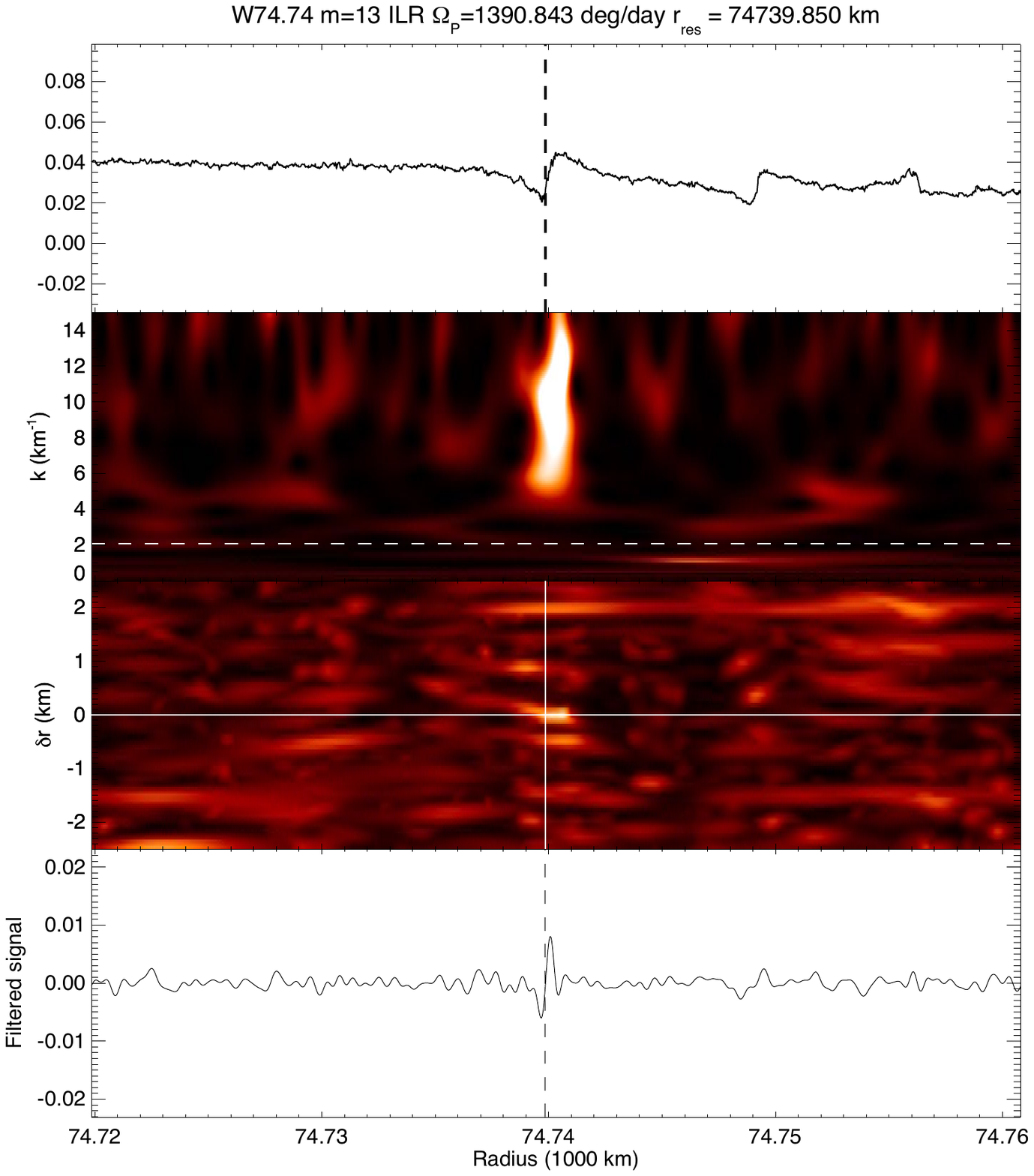}}}
\caption{Phase-corrected wavelet results for the $m= 13$ W74.74 ILR. The top panel shows a distinct feature near 74740 km that is morphologically very similar to the feature just 10 km away, previously identified as an outward-propagating $m=11$ W74.75 ILR. This is a low optical depth region in the inner C ring. The second panel shows strong PCW ratio at high wavenumber outward of the resonance radius. The third panel shows power near $\delta r=0$, indicating a self-consistent pattern speed and resonance radius. The bottom panel shows the reconstructed filtered signal, with a single wave crest outward of the resonance radius.}
\label{fig:W74.74-2D}
\end{figure}

\begin{figure}
{\resizebox{\textwidth}{!}{\includegraphics[angle=0]{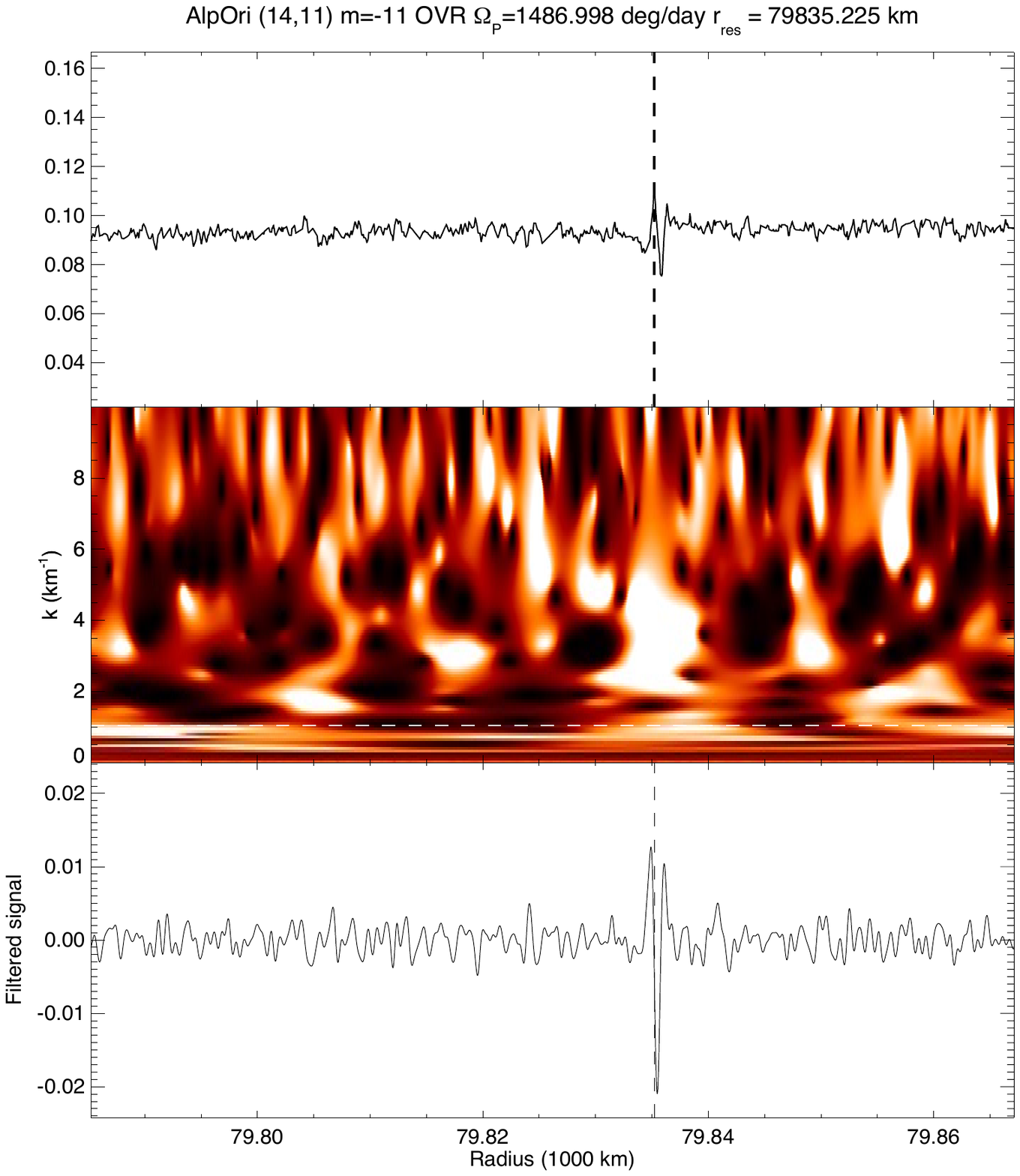}}}
\caption{Phase-corrected wavelet results for the $m= 11$ W79.84 OVR, based on three low-inclination VIMS alpOri occultations (revs 268E, 269E, and 277I) from late in the Cassini mission. The top panel shows a narrow wavelike feature centered on the assumed resonance radius marked by the vertical dashed line. The second panel shows a strong but amorphous PCW ratio signal in the vicinity of the wave. The bottom panel shows the reconstructed filtered signal, with a sharp wave centered on the resonance radius.}
\label{fig:W79.84-2D}
\end{figure}

\begin{figure}
 {\resizebox{\textwidth}{!}{\includegraphics[angle=0]{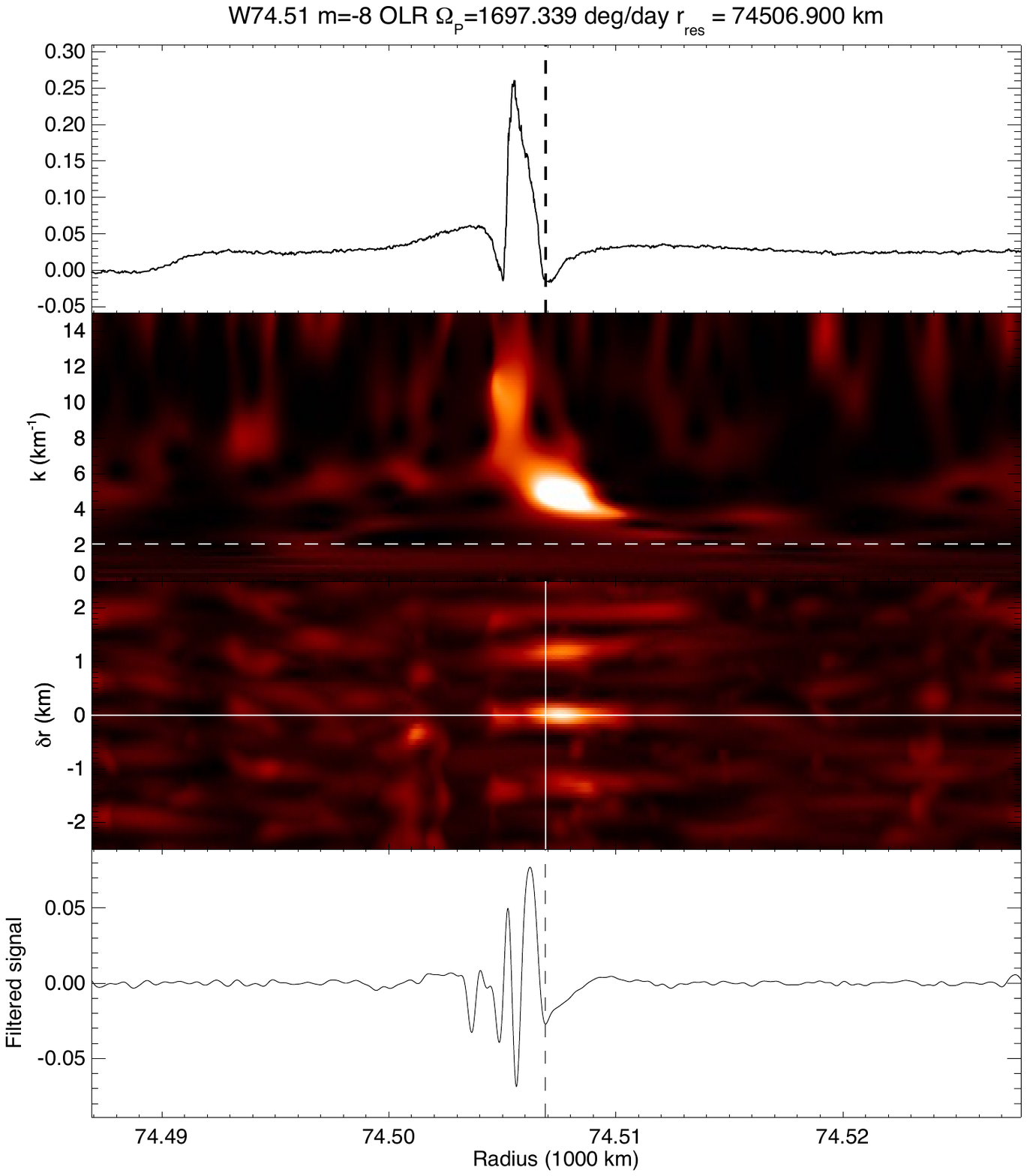}}}
\caption{Phase-corrected wavelet results for the $m= 8$ W74.51 OLR. The top panel shows a prominent feature inward of a narrow gap whose inner edge is coincident with the assumed resonance radius marked by the vertical dashed line. The second panel shows strong PCW ratio at moderate wavenumber outward of the resonance radius and at higher wavenumber inward of the resonance radius. The third panel shows power near $\delta r=0$, indicating a self-consistent pattern speed and resonance radius. The bottom panel shows the reconstructed filtered signal, with a convincing wave inward of the resonance radius.}
\label{fig:W74.51-2D}
\end{figure}

\begin{figure}
{\resizebox{\textwidth}{!}{\includegraphics[angle=0]{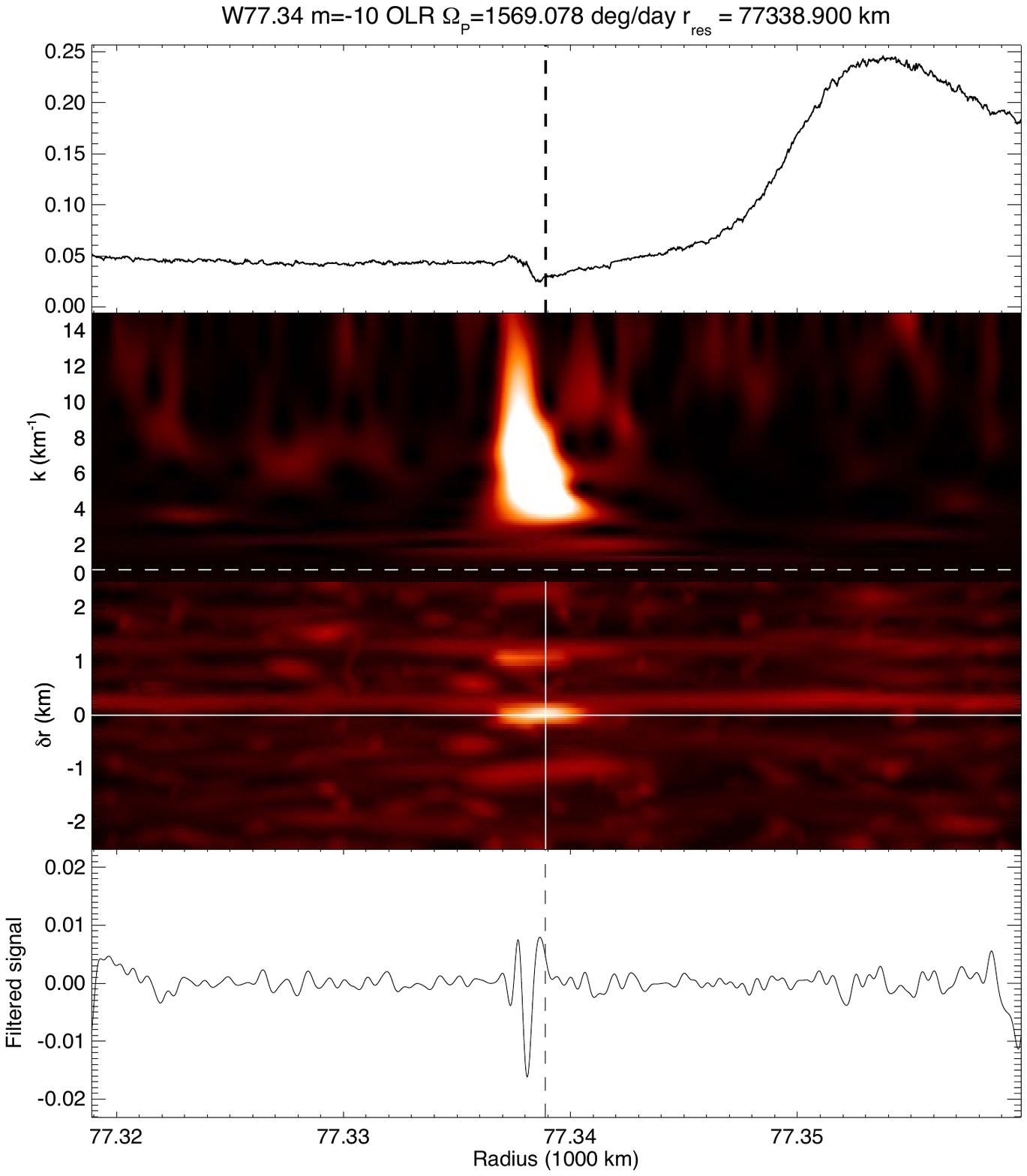}}}
\caption{Phase-corrected wavelet results for the $m= 10$ W77.34 OLR. The top panel shows a distinct feature near 77339 km that is morphologically similar to two other strong OLR detections (W75.14, W79.55). This is a low optical depth region in the inner C ring. The second panel shows strong PCW ratio at moderate wavenumber inward of the resonance radius. The third panel shows power near $\delta r=0$, indicating a self-consistent pattern speed and resonance radius. The bottom panel shows the reconstructed filtered signal, with a convincing wave inward of the resonance radius.}
\label{fig:W77.34-2D}
\end{figure}

\begin{figure}
{\resizebox{\textwidth}{!}{\includegraphics[angle=0]{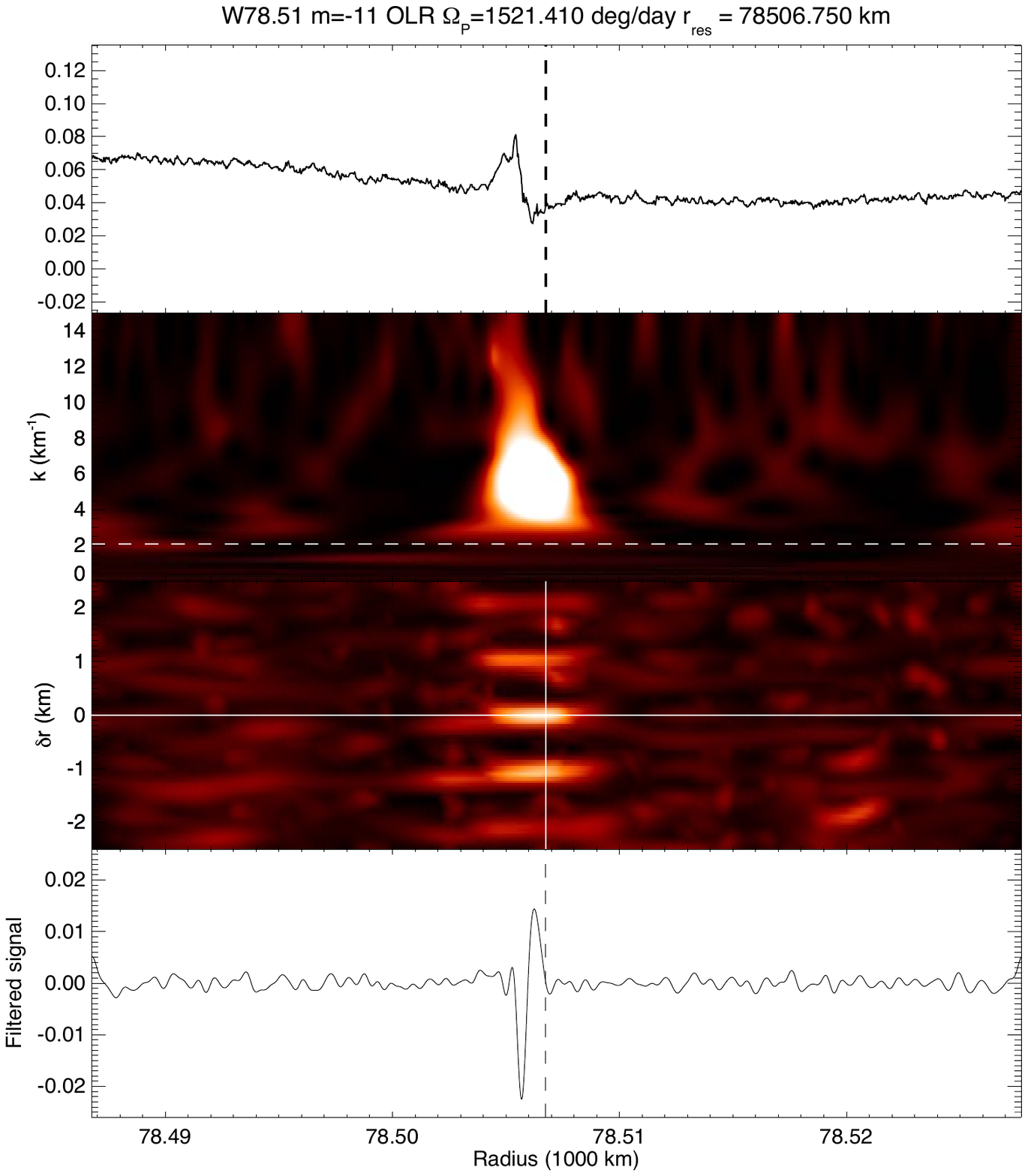}}}
\caption{Phase-corrected wavelet results for the $m= 11$ W78.51 OLR. The top panel shows a distinct feature near 78507 km with the characteristic shape of an unresolved inward-propagating wave. This is a low optical depth region in the inner C ring. The second panel shows strong PCW ratio at moderate wavenumber inward of the resonance radius. The third panel shows power near $\delta r=0$, indicating a self-consistent pattern speed and resonance radius. The bottom panel shows the reconstructed filtered signal, with a convincing wave inward of the resonance radius.}
\label{fig:W78.51-2D}
\end{figure}

\begin{figure}
{\resizebox{\textwidth}{!}{\includegraphics[angle=0]{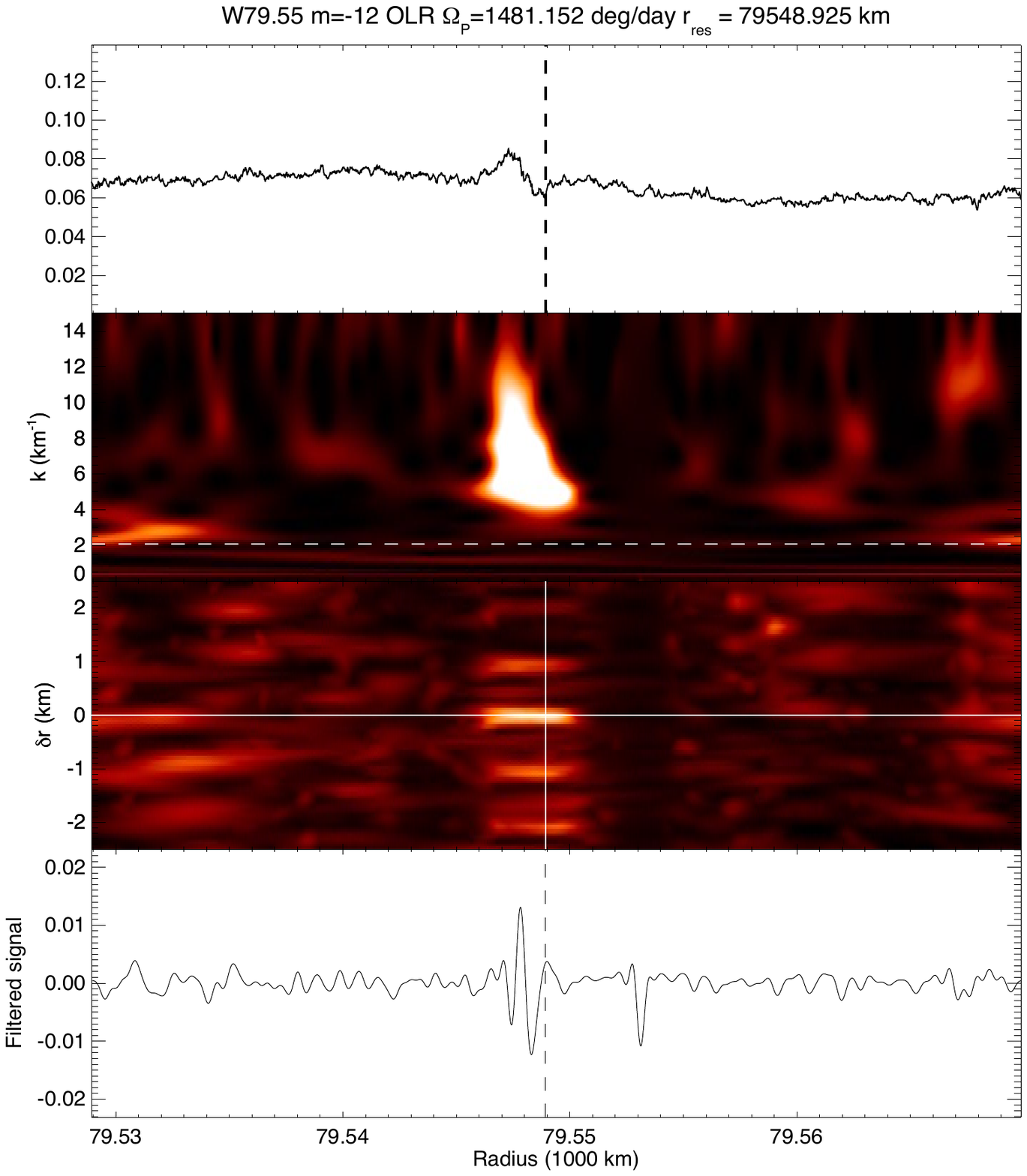}}}
\caption{Phase-corrected wavelet results for the $m= 12$ W79.55 OLR. The top panel shows a distinct feature near 79549 km that is morphologically similar to two other strong OLR detections (W75.14, W77.34). This is a low optical depth region in the inner C ring. The second panel shows strong PCW ratio at moderate wavenumber inward of the resonance radius. The third panel shows power near $\delta r=0$, indicating a self-consistent pattern speed and resonance radius. The bottom panel shows the reconstructed filtered signal, with a convincing wave inward of the resonance radius.}
\label{fig:W79.55-2D}
\end{figure}

\begin{figure}
{\resizebox{5.5in}{!}{\includegraphics[angle=0]{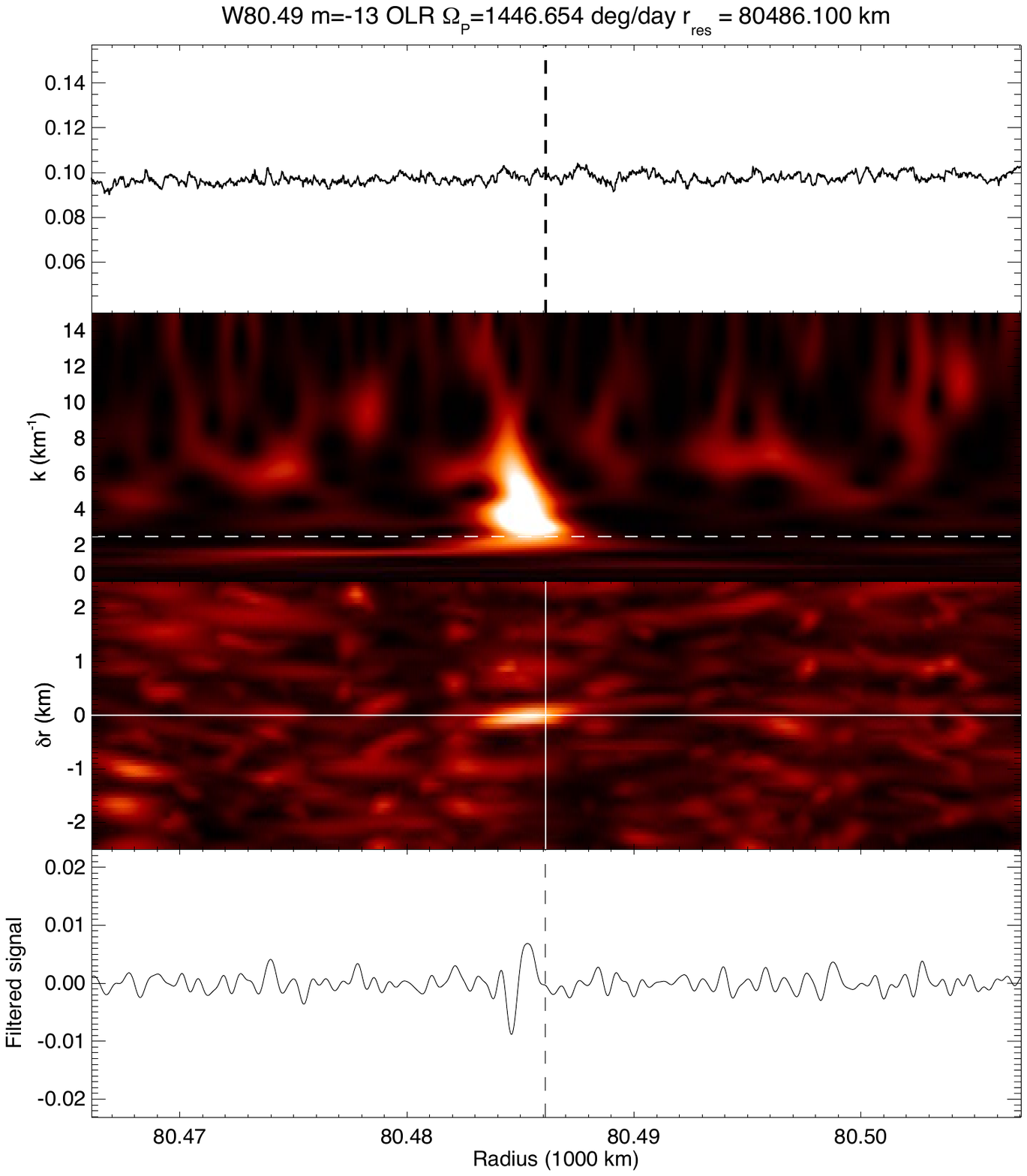}}}
\caption{Phase-corrected wavelet results for the $m=-13$ W80.49 OLR. The top panel shows a nearly featureless region of moderate optical depth, with a bare suggestion of a wave near the resonance radius marked by the vertical dashed line. However, the PCW power ratio spectrum in the second panel shows a strong signal with a negative slope in $k(r)$, inward of the assumed resonance radius. The third panel shows power near $\delta r=0$, indicating that the pattern speed and resonance radius are self-consistent, and the bottom panel shows the reconstructed signal from the PCW, with a convincing wavelike signature inward of the resonance radius, as expected for an OLR.}
\label{fig:W80.49-2D}
\end{figure}

\begin{figure}
{\resizebox{\textwidth}{!}{\includegraphics[angle=0]{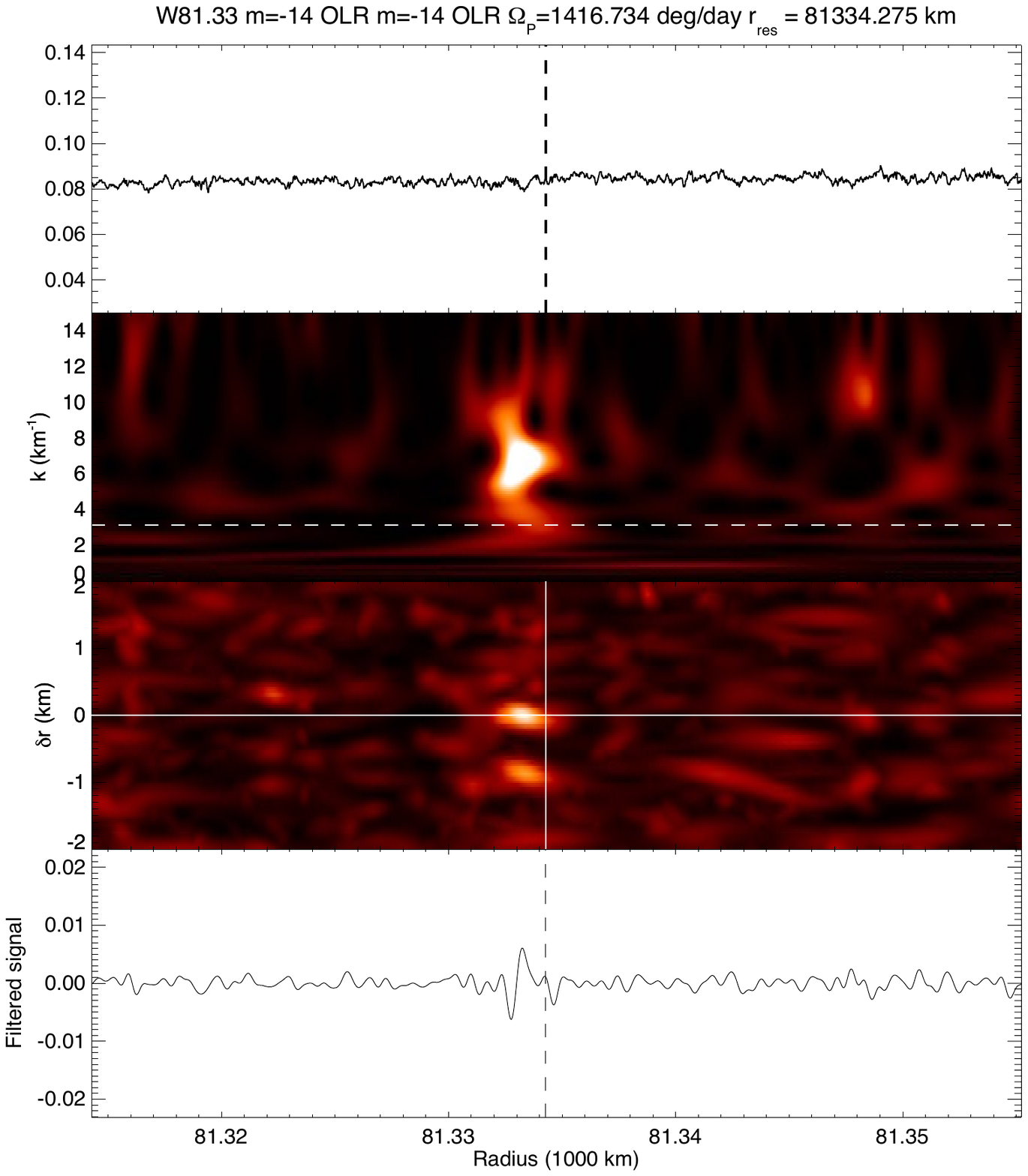}}}
\caption{Phase-corrected wavelet results for the $m= 14$ W81.33 OLR. The top panel shows a nearly featureless region of low optical depth in the middle C ring, with only a hint of a weak wave interior to the resonance radius marked by the dashed vertical line. The second panel shows a rather amorphous PCW ratio at moderate wavenumber inward of the resonance radius. The third panel shows power near $\delta r=0$, indicating a self-consistent pattern speed and resonance radius. The bottom panel shows the reconstructed filtered signal, with a suggestive wave pattern inward of the resonance radius.}
\label{fig:W81.33-2D}
\end{figure}

\begin{figure}
{\resizebox{\textwidth}{!}{\includegraphics[angle=0]{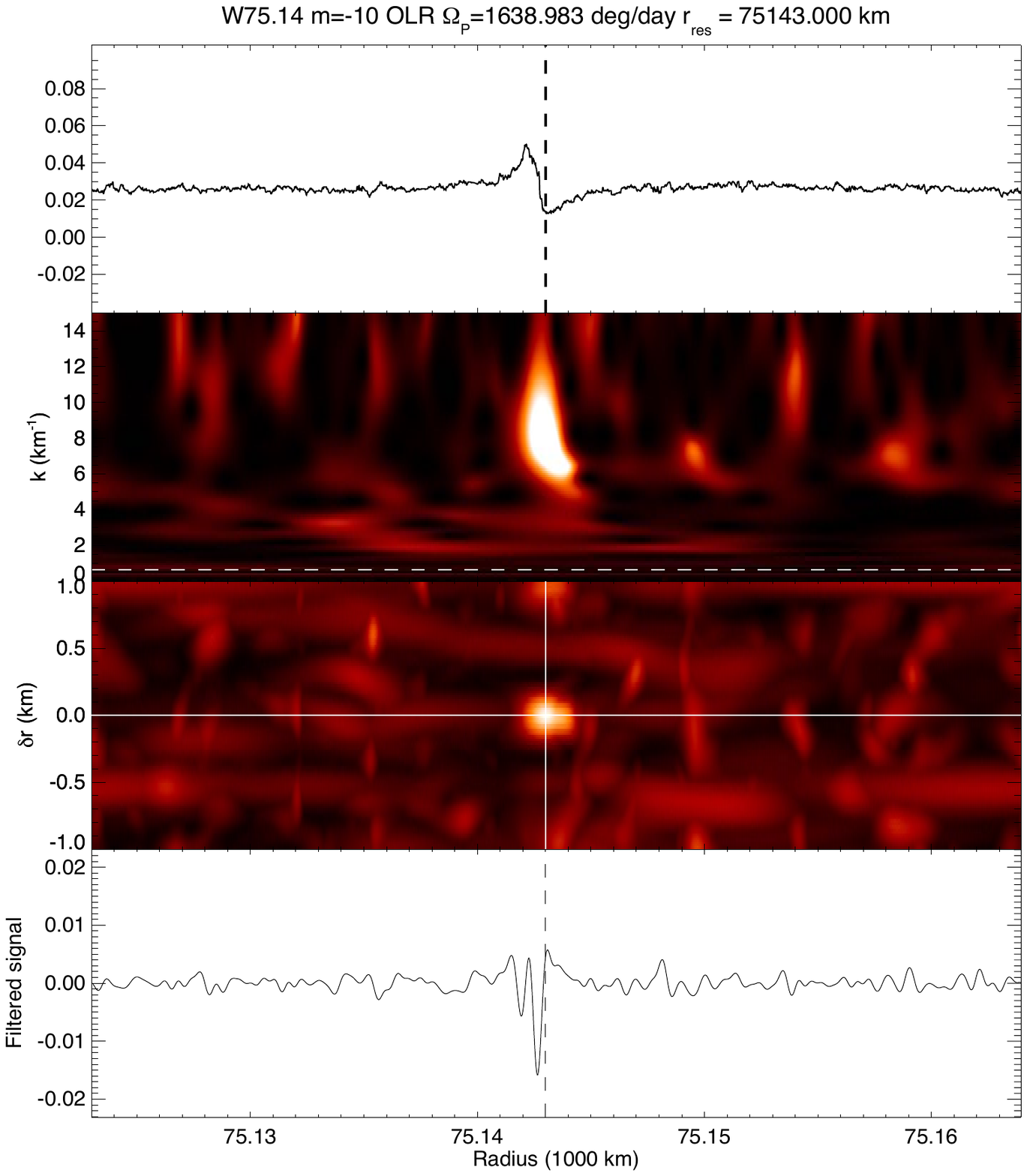}}}
\caption{Phase-corrected wavelet results for the $m= 10$ W75.14 OLR. The top panel shows a distinct feature near 75143 km that is morphologically similar to two other strong OLR detections (W77.34, W79.55). This is a low optical depth region in the inner C ring. The second panel shows strong PCW ratio at moderate wavenumber inward of the resonance radius. The third panel shows power near $\delta r=0$, indicating a self-consistent pattern speed and resonance radius. The bottom panel shows the reconstructed filtered signal, with a convincing wave inward of the resonance radius.}
\label{fig:W75.14-2D}
\end{figure}

\begin{figure}
{\resizebox{5.5in}{!}{\includegraphics[angle=0]{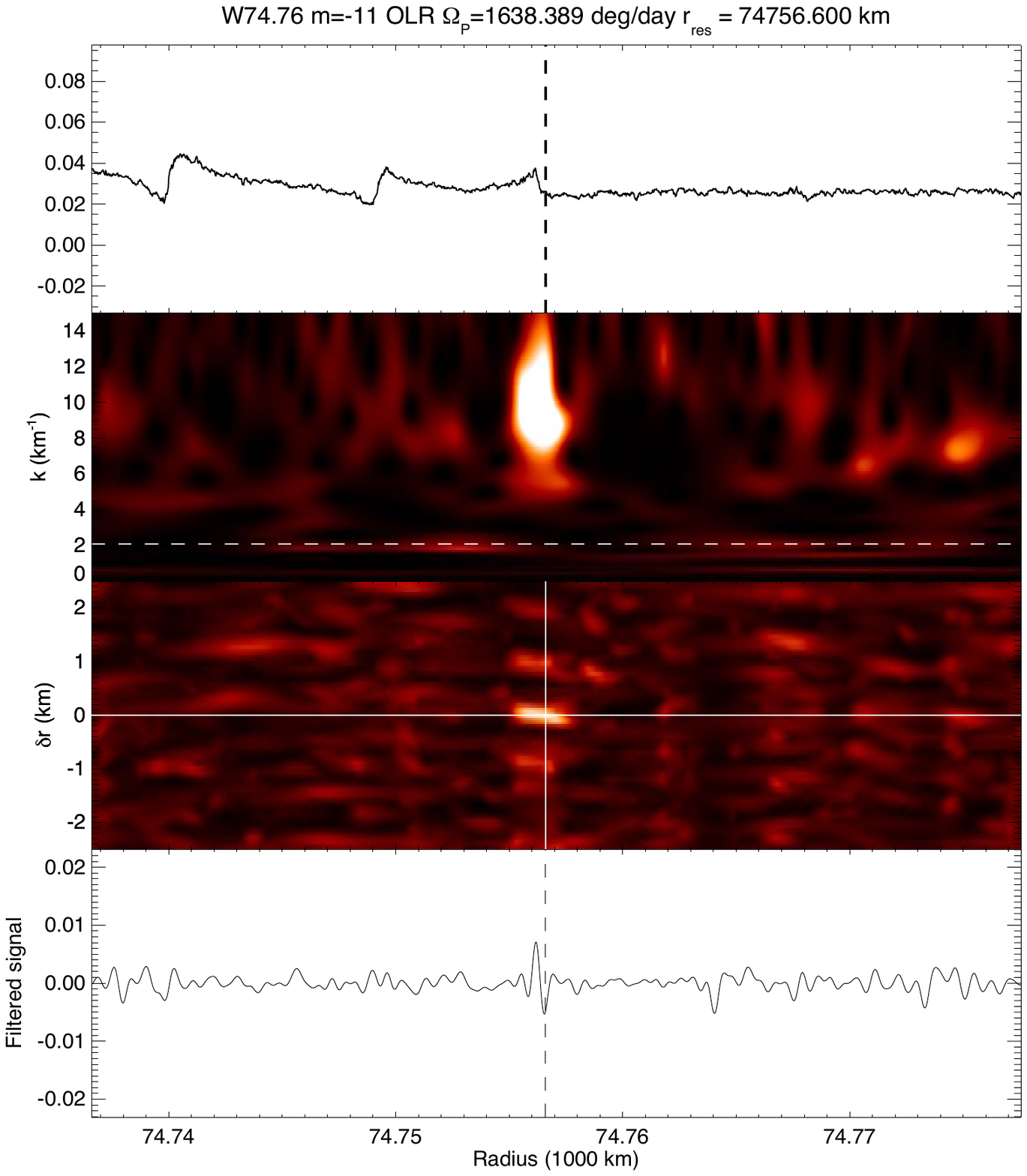}}}
\caption{Phase-corrected wavelet results for the $m= 11$ W74.76 OLR . The top panel shows a distinct feature near 74757 km that is morphologically the mirror reflection of two nearby outward-propagating waves: the $m=11$ W74.75 ILR and the $m=13$ W74.74 ILR. This is a low optical depth region in the inner C ring. The second panel shows strong PCW ratio at high wavenumber inward of the resonance radius. The third panel shows power near $\delta r=0$, indicating a self-consistent pattern speed and resonance radius. The bottom panel shows the reconstructed filtered signal, with a single wave crest inward of the resonance radius.}
\label{fig:W74.76-2D}
\end{figure}

\end{document}